\documentclass[twoside,11pt]{article}

\usepackage[preprint]{jmlr2e}
\usepackage{booktabs}
\usepackage{multirow}
\usepackage{subcaption}
\usepackage{xcolor}
\usepackage{colortbl}
\usepackage{amsmath}
\usepackage{graphicx}
\usepackage{float}
\usepackage{enumitem}

\newcommand{\best}[1]{\cellcolor{gray!50}\textbf{#1}}

\usepackage{lastpage}
\jmlrheading{}{2026}{1-\pageref{LastPage}}{}{}{}{Author List}

\ShortHeadings{Samudra 2}{Yuan et al.}
\firstpageno{1}

\begin{document}

\title{Samudra 2: Scaling Ocean Emulators across Resolutions}

\author{\name Yuan Yuan$^{1}$, Jesse Rusak$^2$, Alexander Merose$^2$, Adam Subel$^1$, Pavel Perezhogin$^1$, Alistair Adcroft$^{3}$, Carlos Fernandez-Granda$^{1}$, Laure Zanna$^{1}$ \\
\addr $^1$Courant Institute School of Mathematics, Computing, and Data Science, New York University, New York, NY, USA, $^2$Open Athena AI Foundation, Inc., New York, NY, USA, $^3$Program in Atmospheric and Oceanic Sciences, Princeton University, Princeton, NJ, USA \\
Corresponding author: yy6080@nyu.edu}

\maketitle

\begin{abstract}
Ocean general circulation models (OGCMs) are essential to climate science but computationally expensive, limiting ensemble size and forcing scenarios. Neural emulators promise orders-of-magnitude speedups, yet existing ocean emulators have not combined fine spatial resolution with multi-year autoregressive rollouts. Samudra, the first autoregressive neural ocean emulator to produce multi-decade global rollouts, is limited to $1^\circ$ resolution and exhibits two long-horizon failure modes: \emph{variance collapse}, the loss of temporal variability, and \emph{imprinting artifacts}, in which velocity patterns leak into deep-ocean fields. We present Samudra 2, which introduces a wider U-Net backbone with modified ConvNeXt-style blocks and a reduced block-internal expansion factor, together with a dynamic loss that reweights output channels according to their prediction errors, strengthening gradients for slow-evolving deep-ocean fields. At $1^\circ$, Samudra 2 increases upper-ocean global-mean temperature $R^2$ from 0.56 to 0.87 and reduces deep-ocean temperature error by roughly sevenfold. The same architecture scales to $1/2^\circ$ and $1/4^\circ$ over approximately 8-year autoregressive rollouts, recovering mesoscale eddies and sharp western boundary currents. Running on a single GPU, Samudra 2 enables larger ensembles for sea-level projections, ocean heat uptake, and climate variability studies.
All artifacts are publicly available:
  \href{https://m2lines.github.io/Samudra/samudra2/}{\color{blue}project page},
  \href{https://github.com/m2lines/Samudra}{\color{blue}code},
  \href{https://huggingface.co/M2LInES/Samudra2}{\color{blue}checkpoints},
  and \href{https://m2lines.github.io/Samudra/docs/}{\color{blue}documentation}.
\end{abstract}


\section{Introduction}
\label{sec:intro}

\begin{figure}[t]
\centering
\includegraphics[width=0.9\linewidth]{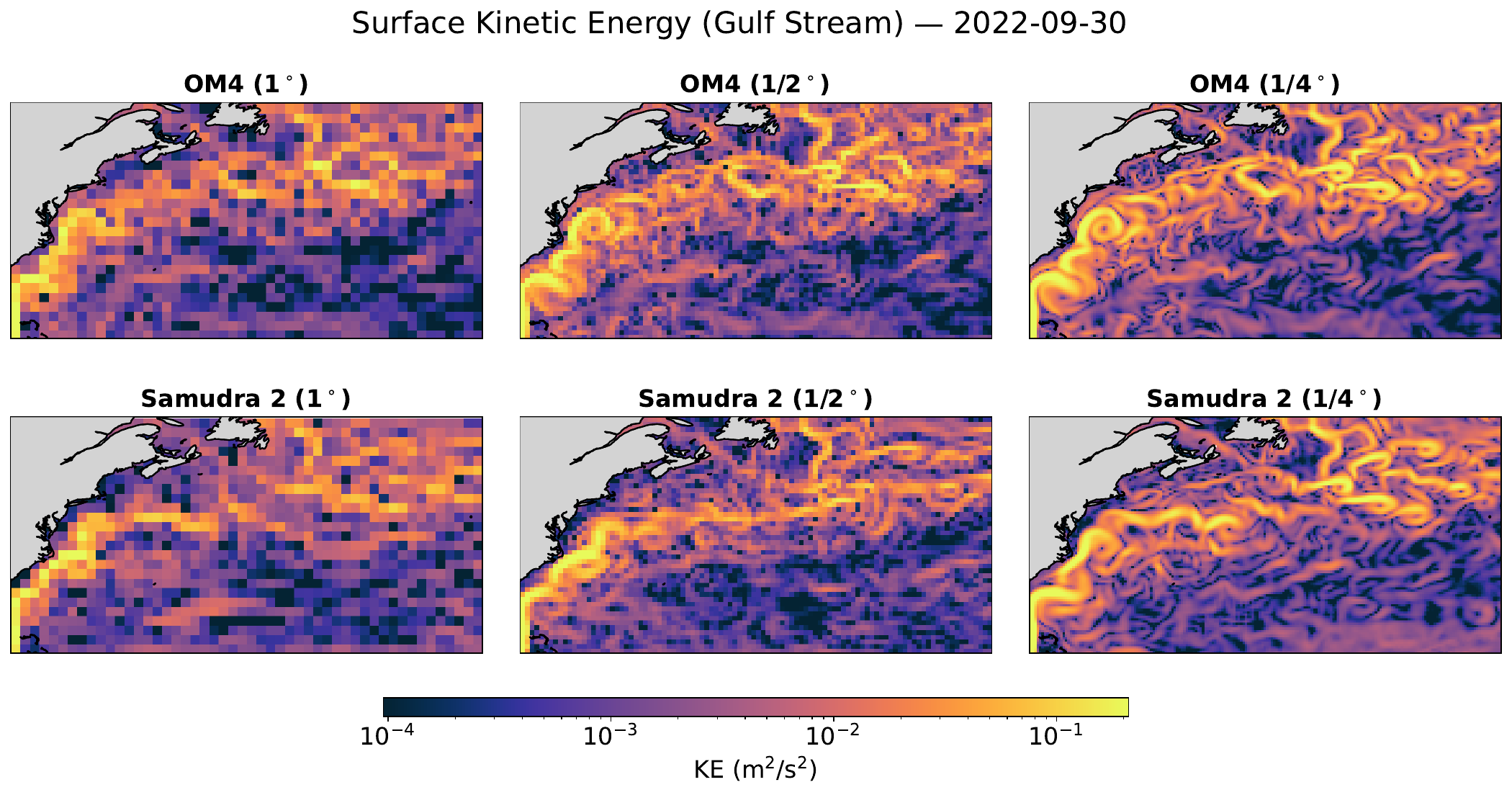}
\caption{Surface kinetic energy in the Gulf Stream region from
GFDL OM4 (top) and our emulator (bottom) at $1^\circ$ (left),
$1/2^\circ$ (middle), and $1/4^\circ$ (right) resolutions.
Snapshots are taken at 2022-09-30, near the end of an 8-year
autoregressive rollout (${\sim}580$ steps). At finer resolutions,
the emulator progressively captures the mesoscale eddies, meanders, and filamentary structures characteristic of the Gulf Stream western boundary current.}
\label{fig:ke_multiscale}
\end{figure}

Ocean general circulation models (OGCMs) are essential to climate science, simulating the global transport of heat, salt, carbon, and momentum that underpins seasonal forecasts, decadal predictions, and centennial climate projections \citep{dunne2024evolving, xu2023enhanced,eyring2016cmip6}.
However, they are computationally expensive: a single century-long eddy-permitting simulation can require millions of core-hours, severely limiting the number of forcing scenarios and ensemble members that can be explored \citep{hewitt2020resolving}.
This expense is driven largely by the need for high spatial resolution: for example, mesoscale eddies and sharp frontal structures that dominate ocean variability only emerge at $1/4^\circ$, while coarser grids produce progressively smoother flow fields (Figure~\ref{fig:ke_multiscale}, top row).
Reducing this cost would enable affordable large ensembles for uncertainty quantification, broader exploration of emission and initial-condition scenarios, and tractable studies of mesoscale- and deep-ocean-dependent processes central to climate projection, such as heat uptake, sea-level rise, and the oceanic carbon sink \citep{dunne2024evolving, duncan2025samudrace, kochkov2024neuralgcm}.

The computational bottleneck of traditional numerical models has
motivated a growing body of work on data-driven \emph{emulators},
primarily for atmospheric forecasting \citep{pathak2022fourcastnet,
bi2023pangu, lam2023graphcast, kochkov2024neuralgcm}, which
leverage machine learning to reproduce the input-output behavior
of physical simulators at a fraction of the cost. For ocean
simulations, Samudra \citep{dheeshjith2025samudra} demonstrated
that autoregressive neural emulation can produce multi-decade
rollouts of key ocean variables at $1^\circ$ resolution. Samudra
was trained on output from GFDL OM4 \citep{adcroft2019gfdl}, a
state-of-the-art ocean general circulation model and the ocean
component of the GFDL CM4 coupled climate model, with related work
spanning transfer learning \citep{dheeshjith2024transfer},
atmosphere-ocean coupling \citep{duncan2025samudrace}, and
high-resolution short-term forecasting \citep{cui2025wenhai}. Yet
no existing ocean emulator combines fine spatial resolution with
the dynamical fidelity and long-term stability for multi-year
climate rollouts.

From a machine learning perspective, ocean emulation poses distinctive challenges.
The goal is to simulate ocean dynamics over the entire planet: given the current high-dimensional ocean state, the model predicts the next state, then the one after that, and so on for years, with each prediction fed back as input.
Training uses short rollouts for computational tractability, but evaluation targets climate-scale rollouts spanning multi-year to decadal horizons, assessed by long-horizon metrics such as the mean state, temporal variability, spectral distribution of variance, and modes of climate variability. Short-horizon training losses do not directly optimize these metrics.
This mismatch is not unique to ocean emulation: related long-horizon failure modes appear as blurring in video prediction \citep{mathieu2016deep}, drift in world models \citep{hafner2023dreamerv3}, and spectral variance loss in atmospheric emulators \citep{lam2023graphcast, kochkov2024neuralgcm, wattmeyer2023ace}, placing our work within a broader class of long-horizon autoregressive ML problems.
In the ocean setting, these challenges concretely limit our direct predecessor, Samudra, in three ways. Two are dynamical manifestations of the train--evaluate mismatch: (i) \emph{variance collapse}, where autoregressive error accumulation pulls predictions toward the climatological mean and suppresses temporal variability, especially at depth; and (ii) \emph{imprinting artifacts}, where velocity-field patterns leak into ocean temperature and salinity patterns, producing spurious banding and high-frequency noise that amplify over long rollouts. The third is a resolution limitation: (iii) Samudra is restricted to coarse ($1^\circ$) resolution, where mesoscale eddies and sharp frontal structures are entirely unresolved \citep{hallberg2013resolution, hewitt2020resolving}.

In this work, we present \textbf{Samudra 2}, which extends Samudra through two complementary modifications.
The first is an \emph{architectural scaling}: we widen the U-Net backbone built from modified ConvNeXt-style blocks \citep{liu2022convnext, ronneberger2015unet} and reduce its block-internal expansion factor, shifting capacity toward inter-stage feature dimensions.
The second is a \emph{dynamic loss function} that reweights each per-channel mean squared error  (MSE) term by the inverse of that channel's running prediction error, adaptively amplifying the gradient signal from slow-evolving deep-ocean fields that standard MSE  would neglect.
The two modifications are complementary by design: the wider backbone provides the representational capacity needed at higher resolution, addressing~(iii), and the dynamic loss is the primary driver of deep-ocean fidelity, addressing~(i) and~(ii).
Figure~\ref{fig:ke_multiscale} (bottom row) previews the resulting multi-scale kinetic-energy structure at all three resolutions.

We evaluate Samudra 2  at three spatial resolutions: $1^\circ$, $1/2^\circ$, and $1/4^\circ$ against the corresponding regridded output of GFDL OM4.
Because single-step accuracy can mask the variance collapse and drift that only emerge over extended rollouts \citep{rasp2024weatherbench}, we assess emulators by long-horizon, climate-relevant diagnostics, including temporal variance, detrended time series, spectral distributions, and indices of climate variability such as the Ni\~no 3.4 index, computed over approximately 8-year autoregressive rollouts.
We evaluate Samudra 2 against OM4 at all three resolutions over multi-year autoregressive rollouts, and additionally compare against the original Samudra at 1$^\circ$.
Ablations at $1^\circ$ further confirm that the wider architecture and the dynamic loss are distinct, complementary, and required for Samudra 2's fidelity and scaling gains.
Our main contributions are as follows:
\begin{itemize}
    \item  \textbf{Samudra 2:} An improved AI ocean emulator that combines a wider ConvNeXt U-Net with a dynamic variance-weighted loss to address variance collapse and imprinting artifacts in long-horizon autoregressive ocean emulation.
    \item \textbf{Scaling ocean emulation to higher resolutions:} A demonstration of multi-year ocean emulation at $1/2^\circ$ and $1/4^\circ$ resolutions on GFDL OM4, showing that higher-resolution emulators recover mesoscale structures inaccessible at coarse resolution.
\end{itemize}

\section{Related Work}
\label{sec:related}

\paragraph{Data-Driven Weather and Climate Modeling.}
Machine learning has emerged as a competitive alternative to physics-based numerical weather prediction.
FourCastNet \citep{pathak2022fourcastnet}, Pangu-Weather \citep{bi2023pangu}, and GraphCast \citep{lam2023graphcast} showed that neural networks trained on reanalysis data can match or exceed operational forecast skill for medium-range atmospheric prediction, while NeuralGCM \citep{kochkov2024neuralgcm} embedded a learned physics module within a differentiable dynamical core for stable multi-year climate simulations.
These systems share an autoregressive paradigm: training uses short-horizon predictions, but inference iteratively unrolls them, introducing a tension between single-step accuracy and long-rollout variability \citep{rasp2024weatherbench}.
Standard MSE training tends to regress predictions toward the climatological mean, a phenomenon termed variance collapse \citep{mathieu2016deep}; mitigation strategies such as multi-step training \citep{lam2023graphcast} and adversarial objectives \citep{ravuri2021skilful} have been explored in the atmospheric setting, but these challenges are amplified in ocean emulation by the seasonal-to-decadal timescales and extreme dynamical range across depth layers.

\paragraph{Ocean Emulation.}
Ocean emulation replaces a full OGCM with a learned surrogate that maps the current ocean state directly to future states, bypassing expensive numerical time-stepping. Samudra \citep{dheeshjith2025samudra} was the first comprehensive demonstration of autoregressive neural ocean emulation, producing multi-decade rollouts of temperature, salinity, velocity, and sea surface height on GFDL OM4 at $1^\circ$ resolution; complementary work addressed transfer learning across CO$_2$ forcing \citep{dheeshjith2024transfer} and coupling with an
atmospheric emulator \citep{duncan2025samudrace}. ORCA-DL \citep{guo2025orcadl} demonstrated global ocean prediction at seasonal to decadal timescales, while WenHai \citep{cui2025wenhai}
achieved eddy-resolving forecasting but only at short horizons ($\sim$10 days) rather than the multi-year rollouts required for climate. Deep learning has also been applied to coupled climate phenomena such as ENSO, with \citet{ham2019deep} and \citet{zhou2023self} showing skillful multi-year El Ni\~no forecasts using direct (non-autoregressive) prediction, complementary to the autoregressive emulation framework studied here.

\paragraph{Ocean Parameterization.}
In contrast to emulation, which replaces the ocean model entirely, parameterization keeps the physics-based model but uses machine learning to represent unresolved subgrid processes that the grid is too coarse to simulate explicitly \citep{foxkemper2019challenges}.
The use of ML to parameterize mesoscale eddies was pioneered by \citet{zanna2020data} through equation discovery, and has since developed along multiple complementary directions across many groups: stochastic-deep learning of momentum forcing \citep{guillaumin2021stochastic}, a posteriori learning in quasi-geostrophic turbulence \citep{frezat2022posteriori}, physics-constrained closures with transfer learning \citep{guan2022stable}, the role of training data and filtering choices \citep{yan2024choice}, online stability and scale-aware closures within MOM6 \citep{perezhogin2024stable}, end-to-end differentiable online learning \citep{maddison2026online}, generalization across configurations via dimensional scaling \citep{perezhogin2025generalizable}, and physics-aware CNNs trained on reanalysis \citep{wang2024applications}.
Beyond mesoscale eddies, ML has been applied to other subgrid processes, including vertical mixing in the ocean surface boundary layer \citep{zhu2022physics} and submesoscale buoyancy fluxes \citep{bodner2025data}.

\section{Method}
\label{sec:method}

\subsection{Ocean Emulation Framework}
\label{sec:framework}

An ocean emulator is an autoregressive deep learning model trained to predict future ocean states from recent ones, replacing the expensive numerical time-stepping of an OGCM with a learned state transition. 
Let $\mathbf{x}_t \in \mathbb{R}^{C \times H \times W}$ denote the ocean state at time step $t$, 
where $C$ is the total number of predicted channels and $H \times W$ the spatial grid dimensions.
Four three-dimensional variables, including potential temperature (\texttt{thetao}), salinity (\texttt{so}), zonal velocity (\texttt{uo}), and meridional velocity (\texttt{vo}), are each discretized on $D$ depth levels, while sea surface height (\texttt{zos}) is a single surface field, giving $C = 4D + 1$ channels.
In addition, the emulator receives atmospheric forcing fields $\mathbf{f}_t$  that drive the ocean from above.

The emulator is a learned function $g_\theta$ that receives two consecutive ocean states and their associated atmospheric forcing fields as input, and predicts the next two states:
\begin{equation}
    (\hat{\mathbf{x}}_{t+1},\, \hat{\mathbf{x}}_{t+2}) = g_\theta(\mathbf{x}_{t-1},\, \mathbf{x}_t,\, \mathbf{f}_{t-1},\, \mathbf{f}_t).
    \label{eq:emulator}
\end{equation}
This 2-in-2-out setup follows the design of Samudra \citep{dheeshjith2025samudra}.

\begin{figure}[t]
\centering
\includegraphics[width=\textwidth]{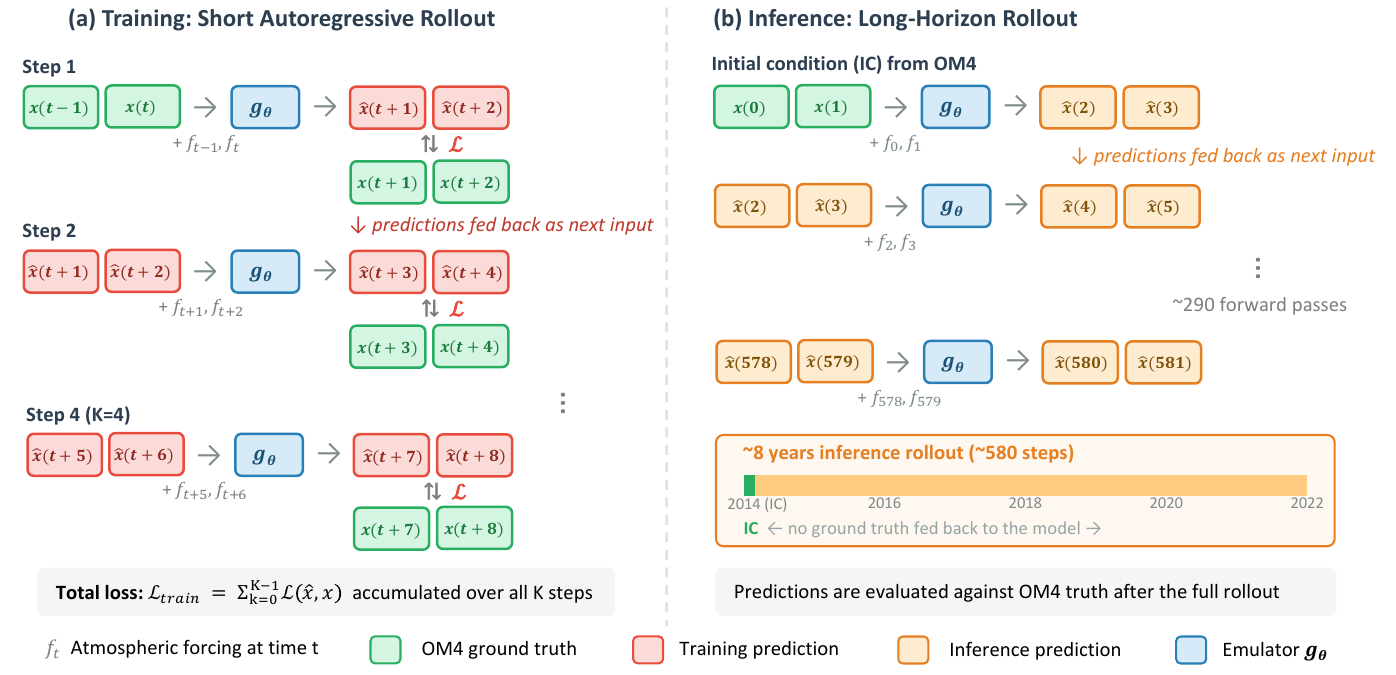}
\caption{Training (a) vs.\ inference (b) rollout. During training, the emulator is unrolled for $K=4$ steps and losses are accumulated against OM4 truth at each step. During inference, it is unrolled $\sim$290 times from a single OM4 initial condition without ground-truth feedback, producing $\sim$580 predicted time steps
($\sim$8 years).}
\label{fig:training_vs_evaluation}
\end{figure}

During training, the model is run autoregressively for a short rollout of $K$ steps: starting from ground-truth states $(\mathbf{x}_{t-1}, \mathbf{x}_t)$, the model predicts the next two time steps, then feeds its own predictions back as input for the next forward pass, repeating $K$ times.
The training loss accumulates over all $K$ steps:
\begin{equation}
    \mathcal{L}_{\text{train}} = \sum_{k=0}^{K-1} \mathcal{L}\!\left(\hat{\mathbf{x}}_{t+2k+1},\, \hat{\mathbf{x}}_{t+2k+2},\, \mathbf{x}_{t+2k+1},\, \mathbf{x}_{t+2k+2}\right),
    \label{eq:train_loss}
\end{equation}
where $\hat{\mathbf{x}}$ denotes predictions and $\mathbf{x}$ ground truth.
We use $K = 4$ autoregressive steps during training (Figure~\ref{fig:training_vs_evaluation}a), so each training sample covers a rollout of 8 predicted time steps (40 days at 5-day resolution).
This \emph{multi-step} training objective exposes the model to its own error accumulation during optimization, providing a stronger training signal for long-horizon stability compared to single-step training.

During evaluation, the emulator is run autoregressively over much longer horizons (Figure~\ref{fig:training_vs_evaluation}b): starting from ground-truth initial conditions $(\mathbf{x}_0, \mathbf{x}_1)$, the model predicts the next two steps, then slides the input window forward by two and repeats.
Defining $\tilde{\mathbf{x}}_0 = \mathbf{x}_0$ and $\tilde{\mathbf{x}}_1 = \mathbf{x}_1$ (ground truth), subsequent states are generated as:
\begin{equation}
    (\tilde{\mathbf{x}}_{t+1},\, \tilde{\mathbf{x}}_{t+2}) = g_\theta(\tilde{\mathbf{x}}_{t-1},\, \tilde{\mathbf{x}}_t,\, \mathbf{f}_{t-1},\, \mathbf{f}_t), \quad t = 1, 3, 5, \ldots
    \label{eq:rollout}
\end{equation}
This \emph{long-horizon} unrolling is the setting that matters for climate emulation, where rollouts may span decades to centuries of simulated time.

\subsection{Data}
\label{sec:data}

We use output from the GFDL OM4p25 ocean model (nominally $1/4^\circ$)  \citep{adcroft2019gfdl}, the ocean component of the GFDL CM4 coupled climate model. We regrid the native tripolar output to Gaussian grids and subsample the vertical coordinate to 19 depth levels. Regridding is performed at three different resolutions:
\begin{itemize}
    \item \textbf{$1^\circ$}: the same resolution used by Samudra.  Mesoscale eddies are entirely unresolved, and their effects are represented by subgrid-scale parameterizations.
    \item \textbf{$1/2^\circ$}: an intermediate resolution at which the largest mesoscale features begin to be resolved, though most remain parameterized.
    \item \textbf{$1/4^\circ$}: eddy-permitting resolution, where western boundary currents and mesoscale eddies are partially but not fully resolved.
\end{itemize}

At each resolution, we extract four three-dimensional prognostic variables at all 19 depth levels—\texttt{thetao}, \texttt{so}, \texttt{uo}, and \texttt{vo}—plus \texttt{zos} as a single surface field, yielding $4 \times 19 + 1 = 77$ prognostic channels.
The atmospheric forcing consists of four boundary fields: zonal wind stress (\texttt{tauuo}), meridional wind stress (\texttt{tauvo}), net surface heat flux (\texttt{hfds}), and heat flux anomalies (\texttt{hfds\_anomalies}).
All fields are stored as 5-day averages in Zarr format for efficient I/O.
Training spans 1975--2013; inference rollouts run from 2014 to 2022.

\subsection{Architecture: Wider U-Net with Modified ConvNeXt-Style Blocks}
\label{sec:arch}

Samudra 2 builds on the U-Net architecture introduced in Samudra. Specifically, Samudra uses an encoder-decoder architecture with skip connections and modified ConvNeXt-style residual blocks \citep{liu2022convnext}. Unlike the canonical ConvNeXt block, which uses a $7 \times 7$ depthwise convolution, our implementation uses standard 2D convolutions with a smaller kernel size, together with dilation and upscaling choices inherited from Samudra.
We therefore refer to this module as a modified ConvNeXt-style block rather than a canonical ConvNeXt block.
The architectural modifications in Samudra 2 are twofold. First, the channel widths throughout the U-Net are increased from $[200, 250, 300, 400]$ to $[280, 380, 480, 520]$, with corresponding increases in the decoder, providing greater capacity to represent multi-scale spatial patterns. 
This widening over deepening choice follows established design principles in deep learning \citep{zagoruyko2016wide} and recent findings in data-driven weather forecasting \citep{siddiqui2024exploring}.
Second, the internal expansion factor within each modified ConvNeXt-style block is reduced from 4 to 2. This design balances the increased representational capacity of the wider backbone against computational cost, while shifting parameters from the block-internal bottleneck to the inter-stage feature dimensions.

\subsection{Dynamic Loss Function}
\label{sec:dynamic_loss}

A standard MSE training objective treats all variables and depth levels equally:
\begin{equation}
    \mathcal{L}_{\text{MSE}} = \frac{1}{N} \sum_{v=1}^{V} \sum_{d=1}^{D} \sum_{i,j} \left( \hat{x}_{v,d,i,j} - x_{v,d,i,j} \right)^2,
    \label{eq:mse}
\end{equation}
where $N$ is the total number of elements summed over.
In practice, this objective is dominated by channels with high spatial variability, typically surface temperatures and velocities, because global normalization yields large values in energetically active regions, which disproportionately contribute to the loss.
In contrast, deep-ocean variables, which vary slowly in time and exhibit little spatial heterogeneity relative to their global statistics, contribute negligibly to the loss.
This imbalance contributes to the imprinting artifacts observed in Samudra: the optimizer neglects deep-ocean signals because their contribution to the loss is dominated by surface variability.

To address this, we introduce a \emph{dynamic loss function} that reweights each output channel so that no single channel dominates the gradient signal, regardless of its absolute error magnitude.
A wide range of predefined ad hoc scalings have been used throughout the literature, with choices such as channel standard deviation \citep{watt2025ace2, lam2023graphcast}, variance of the time-step residuals \citep{keisler2022forecasting, kochkov2024neuralgcm}, or estimates of MSE from a model earlier in development \citep{bi2023pangu}. 
Our scheme differs in that per-channel weights are updated online from the model's evolving prediction error, rather than fixed a priori from target-data statistics, allowing the loss to adaptively amplify channels the model is underfitting (such as slow-evolving deep-ocean fields).

\begin{equation}
    \mathcal{L}_{\text{dynamic}} = \sum_{t} \sum_{c} \lambda(n, c) \left( \hat{x}_{t,c} - x_{t,c} \right)^2,
    \label{eq:dynamic_loss}
\end{equation}
where $c$ indexes the output channels (each variable-depth combination, plus sea surface height), and $\lambda(n, c)$ is a per-channel weight at training iteration $n$.
The intuition is to set these weights inversely proportional to each channel's prediction error, so that channels with small errors (slow-evolving deep-ocean fields) receive larger weights that amplify their gradient signal, while channels with large errors (energetic surface fields) receive smaller weights to prevent them from dominating the loss. However, unconstrained inverse weighting risks over-amplifying channels whose errors are small simply because they carry little learnable signal, potentially degrading high-variance fields that drive most of the dynamics. The weights must therefore be clamped to balance these objectives (see below).

Since the weights depend on the model's evolving prediction errors, they cannot be determined a priori.
We therefore estimate them online: between batches, the weights are updated using an exponential moving average (EMA) of the inverse per-channel MSE:
\begin{equation}
    \lambda(n, c) = \frac{1}{W}\left((W - 1)\,\lambda(n-1, c) + \dfrac{1}{\sum_{h,w} \left( \hat{x}_{1,c}^{(n)} - x_{1,c}^{(n)} \right)^2}\right),
    \label{eq:lambda_update}
\end{equation}
where $n$ indexes the training iteration, $W$ is the EMA window size, and $\hat{x}_{1,c}^{(n)}$ denotes the single-step prediction at the current iteration.
This update rule smoothly tracks the inverse prediction error for each channel: channels with small current errors maintain larger weights, thereby progressively increasing the gradient signal from channels that standard MSE would otherwise neglect.
The weights are initialized uniformly ($\lambda(0, c) = 1$ for all $c$) and clamped so that $\lambda_{\max} \leq L \cdot \lambda_{\min}$ at each update, 
where $L$ is a configurable cap that controls the maximum relative emphasis placed on low-error channels. We use $L=20$ in all experiments.

\subsection{Training at Different Resolutions}
\label{sec:multireso}

We train three independent models on OM4 data coarse-grained to $1^\circ$, $1/2^\circ$, and $1/4^\circ$ resolutions, using the same architecture and hyperparameters aside from minor adaptations required by memory constraints at $1/4^\circ$ resolution (see Appendix~\ref{sec:impl_details} for details).
Each resolution model is trained from scratch on the corresponding OM4 data.
Moving from $1^\circ$ to $1/4^\circ$ increases the number of grid points by a factor of ${\sim}16\times$, which proportionally increases the memory and compute requirements for each training step. 
Improving the speed of loading data to GPU (primarily through parallelism and careful profiling  to identify and eliminate bottlenecks) and reducing peak memory usage (primarily through  gradient checkpointing) were used to keep training tractable on the available hardware.

All three resolution models are trained for 70 epochs with the Adam optimizer \citep{kingma2015adam}, a learning rate of $6 \times 10^{-4}$, cosine annealing, and an effective batch size of 32 across 8 GPUs. An exponential moving average (EMA) of the parameters is maintained for evaluation.
We do not use early stopping; the final model is the last-epoch checkpoint (see Appendix~\ref{sec:impl_details} for rationale).
Full training details are provided in Appendix~\ref{sec:impl_details}.

\begin{figure}[t]
\centering
\includegraphics[width=0.75\textwidth]{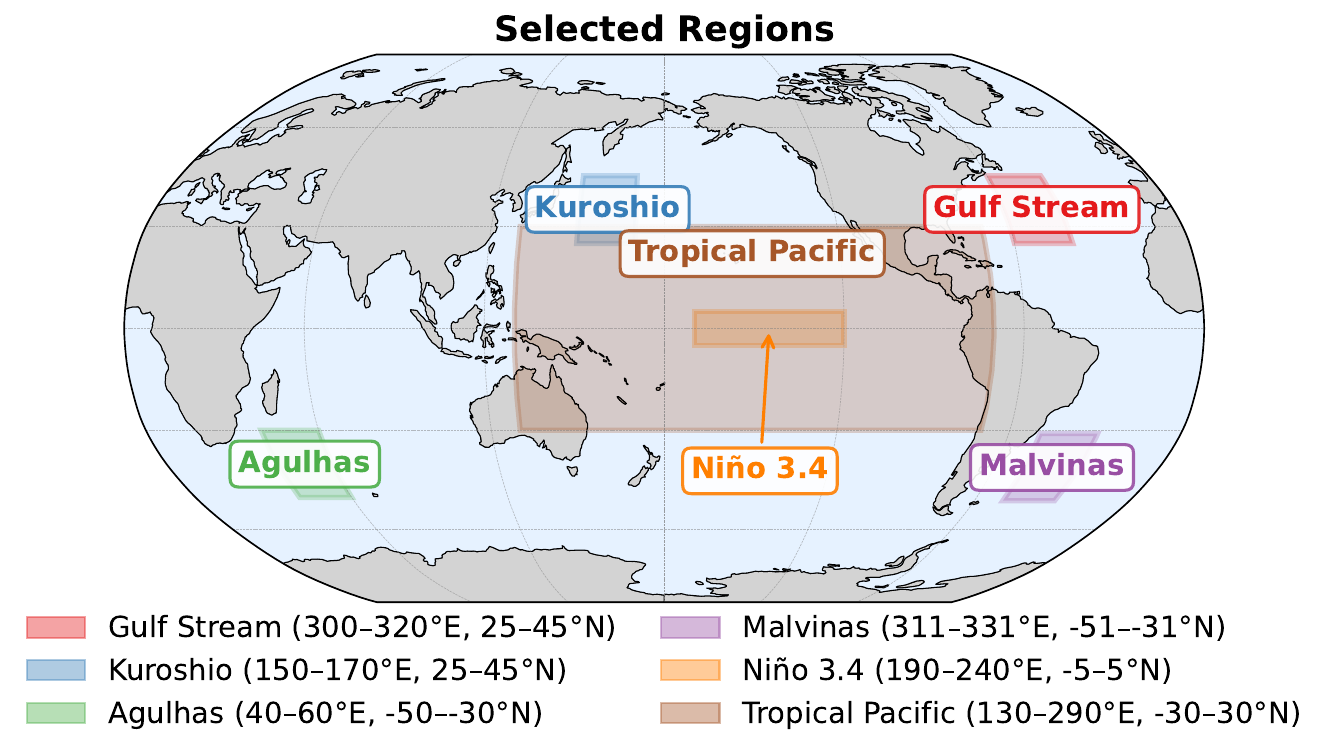}
\caption{The six evaluation regions: four western boundary current regions (Gulf Stream, Kuroshio, Agulhas, Malvinas) and two tropical regions (Ni\~no 3.4, Tropical Pacific). Coordinate bounds are shown in the legend.}
\label{fig:region_map}
\end{figure}

\subsection{Evaluation}
\label{sec:evaluation}

We distinguish between short-horizon training metrics, which measure short-term prediction accuracy, and long-horizon evaluation metrics, which assess the emulator's fidelity over multi-year rollouts using climate-relevant diagnostics.
Many diagnostics below are computed on deseasonalized anomalies: the field minus its pentad-of-year climatology over the training period (1975--2013), which isolates interannual and longer-period variability from the dominant annual cycle.
Short-horizon metrics include per-step RMSE and Pearson correlation.
Long-horizon metrics include temporal variance, detrended global mean
temperature time series, spectral distributions, and indices of climate variability, such as the Ni\~no 3.4 index.
Long-horizon metrics are essential for assessing ocean emulators intended for climate applications, since short-term accuracy can mask variance collapse and drift that only emerge over extended rollouts.
Where spatial snapshots are shown, we select a time step near the end of the rollout (2022-09-30, step ${\sim}575$ of ${\sim}580$), as this represents the stage of maximum autoregressive error accumulation and thus provides the most stringent test of emulator fidelity.
Full mathematical definitions of all metrics are provided in Appendix~\ref{sec:impl_details}.

In addition to global metrics, we evaluate regional performance across six ocean regions (Figure~\ref{fig:region_map}), including four western boundary current (WBC) regions, namely the Gulf Stream, Kuroshio, Agulhas, and Malvinas, where intense mesoscale eddy activity produces some of the strongest temporal variability, and two tropical regions, Ni\~no 3.4 and the Tropical Pacific, which capture large-scale climate modes.
\section{Results}
\label{sec:results}

We first evaluate Samudra 2 against the original Samudra at $1^\circ$ resolution, showing that it maintains and improves performance on key diagnostics (Section~\ref{sec:v1_vs_v2}), and diagnose the imprinting failure mode that limited Samudra's deep-ocean fidelity (Section~\ref{sec:imprinting}). We then scale Samudra 2 to $1/2^\circ$ and $1/4^\circ$ resolutions, examining how spatial, temporal, and spectral fidelity change with resolution (Sections~\ref{sec:variance}--\ref{sec:spectral_results}). Finally, ablation studies isolate the contributions of the wider architecture and dynamic loss (Section~\ref{sec:ablation}).
All models are assessed over autoregressive rollouts of $\sim$580 steps(290 passes, $\sim$8 years) from the same initial condition in the test period, against the OM4 truth at the corresponding resolution.

\begin{figure}[t]
\centering
\includegraphics[width=\textwidth]{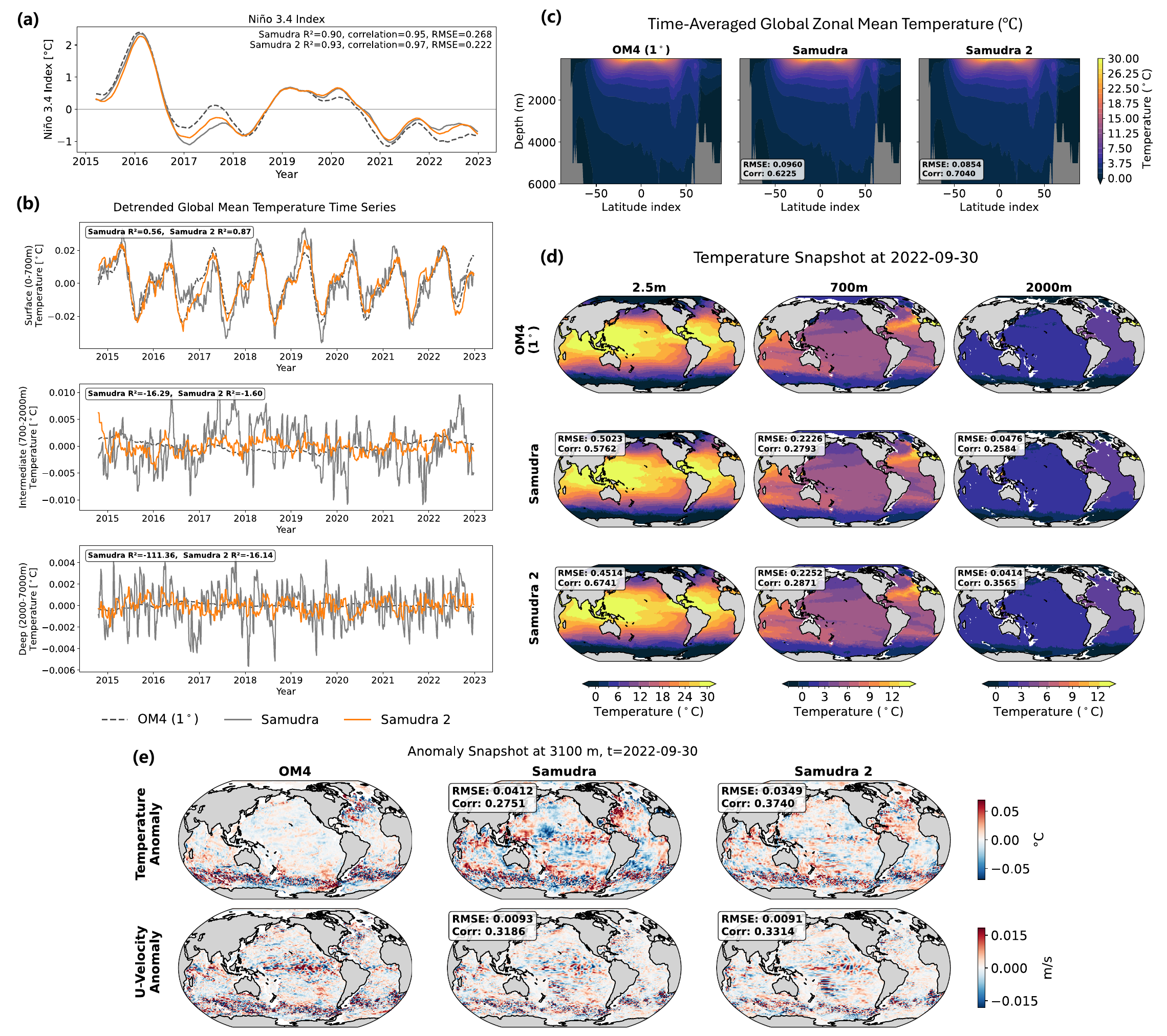}
\caption{Overview comparison of Samudra and Samudra 2 at $1^\circ$ resolution against OM4.
(a) Ni\~no 3.4 index time series, with $R^2$, correlation, and RMSE annotated; dashed, OM4; grey, Samudra; orange, Samudra 2.
(b) Detrended global mean temperature time series for the upper (0--700\,m), intermediate (700--2000\,m), and deep ocean (2000--7000\,m), with $R^2$ values annotated. Note that y-axis scales differ across depth ranges.
(c) Time-averaged zonal mean temperature cross-sections.
(d) Temperature snapshots at 2022-09-30, near the end of the 8-year rollout, at 2.5\,m, 700\,m, and 2000\,m.
For (c) and (d), fields show raw temperature, while annotated RMSE and correlation are computed on deseasonalized anomalies.
(e) deseasonalized temperature anomaly (top) and zonal velocity anomaly (bottom) at 3100\,m for OM4, Samudra, and Samudra 2.}
\label{fig:overview_v1_v2}

\end{figure}

\subsection{Improving upon the Original Samudra}
\label{sec:v1_vs_v2}

Figure~\ref{fig:overview_v1_v2} provides a five-panel overview comparing Samudra and Samudra 2 at $1^\circ$ resolution against the OM4 truth.
Panel~(a) shows the Ni\~no 3.4 index: both models track OM4 well, but Samudra 2 achieves a higher $R^2$ (0.93 vs.\ 0.90) and lower RMSE (0.222 vs.\ 0.268\,$^\circ$C).
Panel~(b) shows detrended global mean temperature time series at three depth ranges.
In the upper ocean (0--700\,m), Samudra 2 improves $R^2$ from 0.56 to 0.87, closely tracking interannual variability while Samudra exhibits substantial high-frequency noise.
At intermediate and deep levels, both models yield negative $R^2$ (intermediate: $-16.29 \to -1.60$; deep: $-111.36 \to -16.14$), meaning predictions at these depths remain worse than a simple temporal mean baseline.
Samudra 2 substantially reduces the magnitude of these errors (by roughly $10\times$ at intermediate depth and ${\sim}7\times$ at deep levels), primarily by suppressing spurious variance from imprinting artifacts (see Section~\ref{sec:imprinting}), but the persistently negative $R^2$ indicates that faithfully tracking deep-ocean variability over multi-year rollouts remains an open challenge.
The root difficulty is that deep-ocean temperature anomalies are
extremely small (order $10^{-3}\,^\circ$C) relative to the variance
of the background ocean state, so even modest spurious fluctuations
dominate the signal and inflate the squared-error numerator of $R^2$.
Panels~(c) and~(d) display raw temperature fields, with RMSE and correlation computed on deseasonalized anomalies.
Panel~(c) shows time-averaged zonal mean cross-sections: Samudra 2 improves the deseasonalized correlation from 0.62 to 0.70 and reduces RMSE from 0.096 to 0.085\,$^\circ$C.
Panel~(d) shows a temperature snapshot at 2022-09-30 at three depths: Samudra 2 improves the deseasonalized correlation at all depths (e.g., 2.5\,m: 0.67 vs.\ 0.58; 700\,m: 0.28 vs.\ 0.27; 2000\,m: 0.36 vs.\ 0.26), though absolute correlations remain moderate after $\sim$8 years of rollout.
Together, these diagnostics confirm that Samudra 2 reproduces the key characteristics of the original Samudra while substantially improving upper-ocean fidelity and reducing deep-ocean errors.

\subsection{Imprinting Artifacts}
\label{sec:imprinting}

\begin{figure}[t]
\centering
\includegraphics[width=0.9\textwidth]{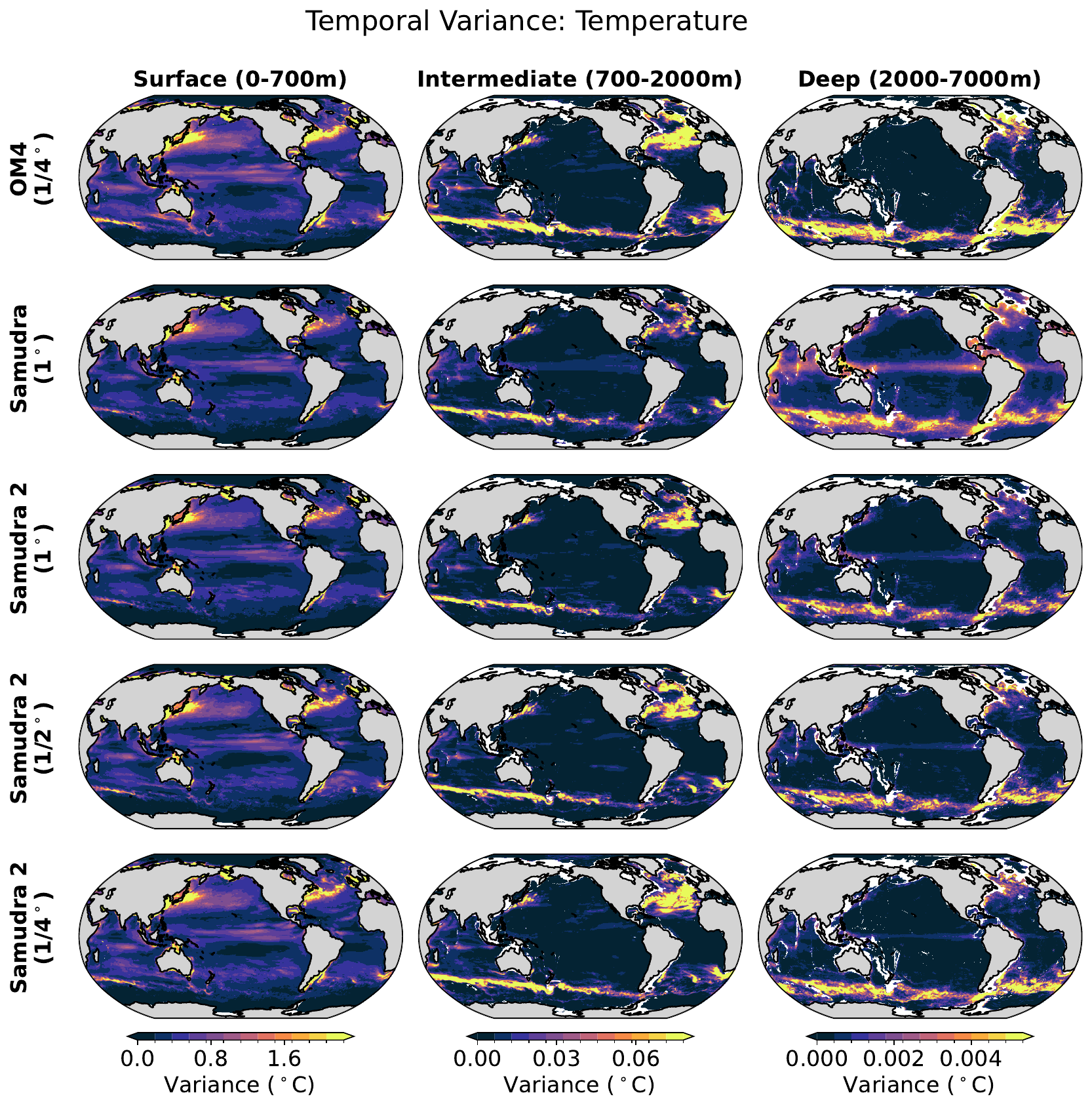}
\caption{Temporal variance of temperature across depth layers (columns) and models (rows). Samudra captures upper-ocean WBC variance but misses deeper hotspots and exhibits imprinting at depth. Samudra 2 recovers realistic variance patterns across resolutions, with sharper WBC hotspots at higher resolution.}
\label{fig:variance_thetao}

\end{figure}

\begin{figure}[t]
\centering
\includegraphics[width=0.9\textwidth]{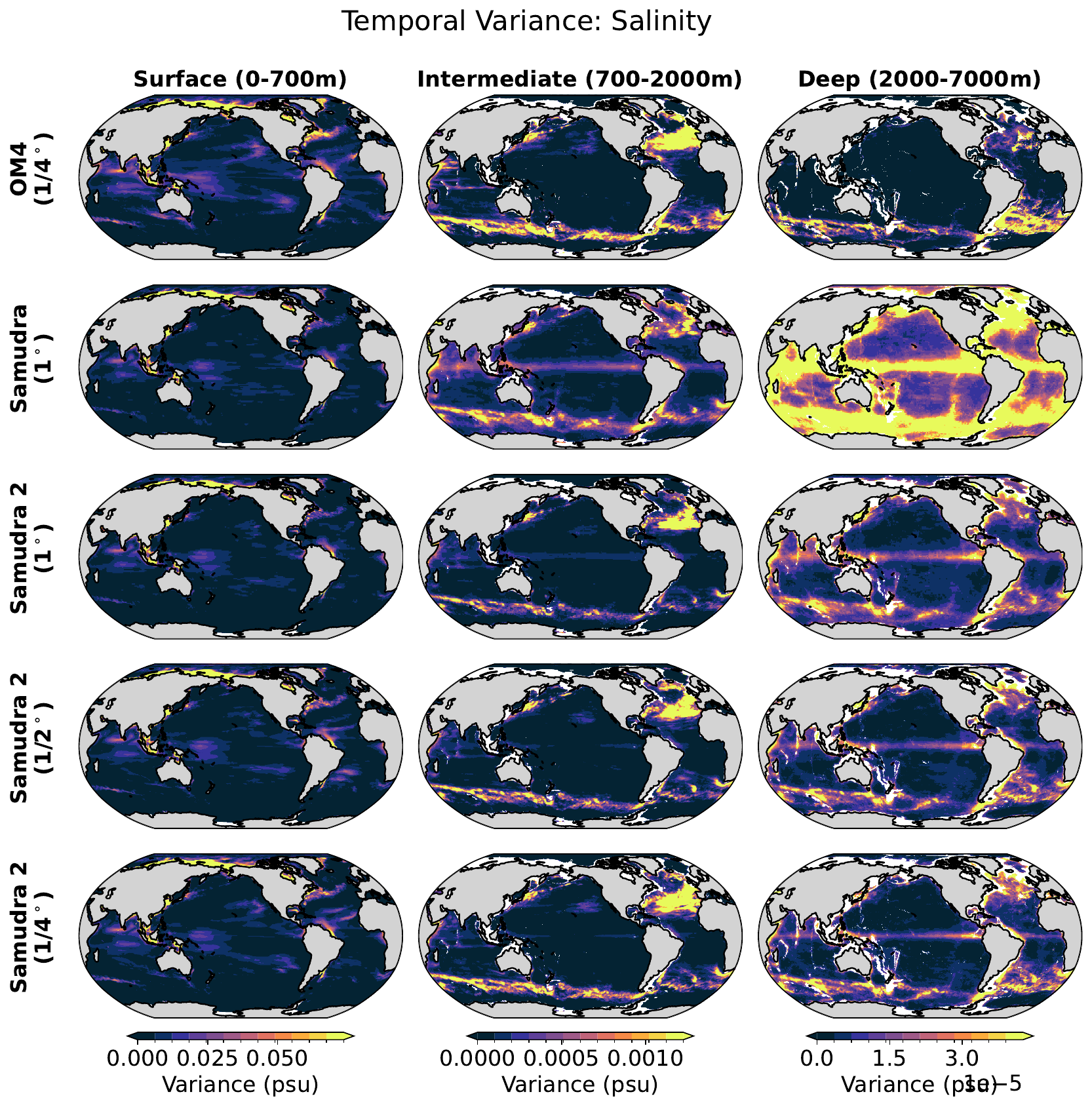}
\caption{Temporal variance of salinity across depth layers and models. Samudra shows widespread spurious deep-ocean variance, whereas Samudra 2 largely suppresses these artifacts and produces variance fields much closer to OM4 across depths.}
\label{fig:variance_so}

\end{figure}

The high-frequency fluctuations in the detrended deep-ocean temperature time series in Figure~\ref{fig:overview_v1_v2}(b) reflect a specific failure mode we term imprinting: velocity field spatial patterns leak into deep-ocean temperature and salinity predictions. This produces nonphysical structure and variability in fields that should be nearly quiescent at depth.
The name reflects the mechanism, as the zonal velocity structure is literally ``imprinted'' onto unrelated variables. This artifact manifests in two ways: (i) deep-ocean variance maps exhibit nonphysical spatial structure, including horizontal banding in temperature (Figure~\ref{fig:variance_thetao}) that mirrors the zonal velocity pattern and widespread elevated variance in salinity (Figure~\ref{fig:variance_so}); and (ii) the detrended global mean temperature time series in Figure~\ref{fig:overview_v1_v2}(b) displays high-frequency fluctuations absent in the OM4 truth, indicating spurious temporal noise that survives global averaging.

\begin{figure}[t]
\centering
\includegraphics[width=0.9\textwidth]{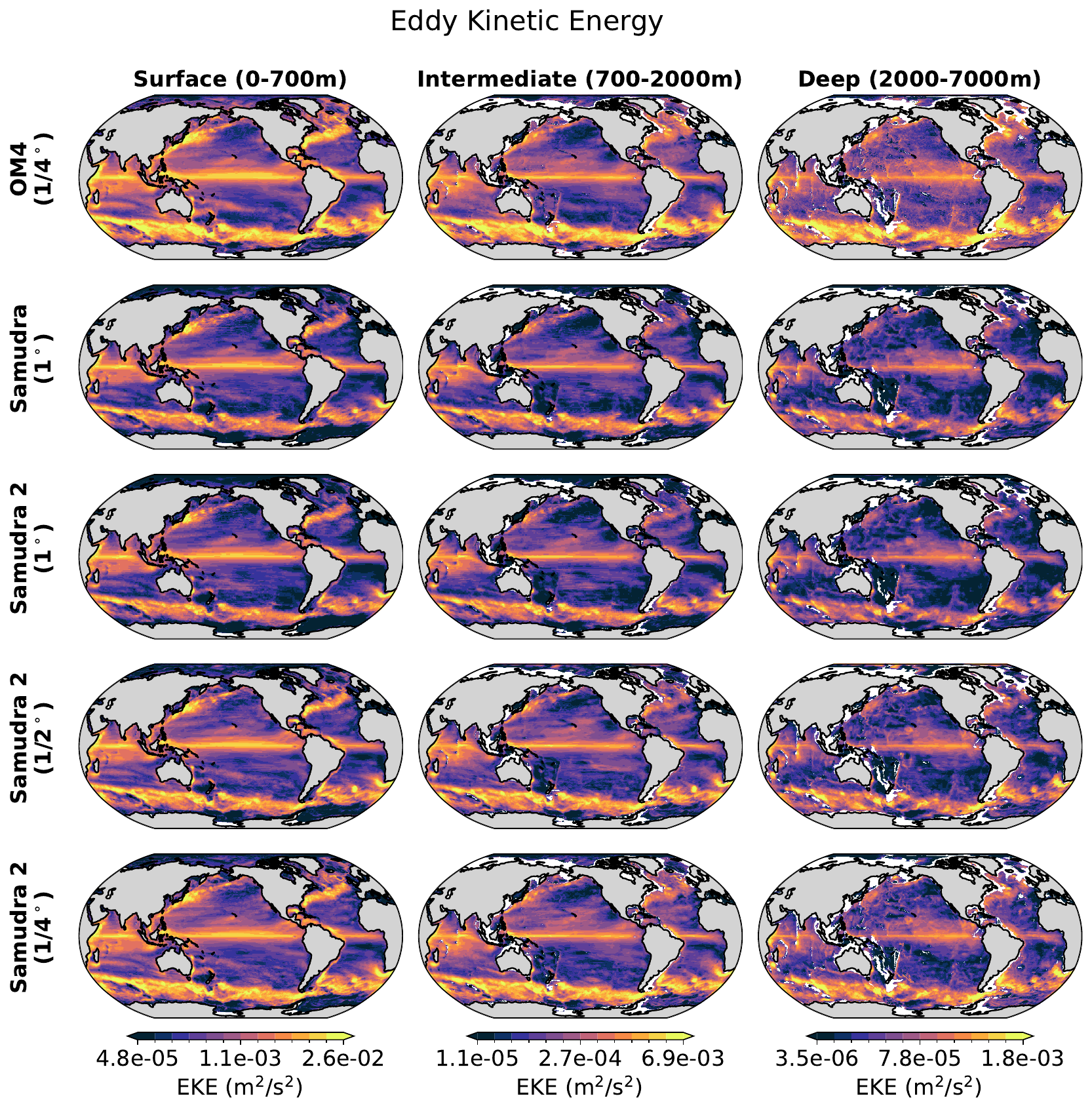}
\caption{Temporal variance of EKE across depth layers (columns) and models (rows). Logarithmic color scale. 
Across Samudra 2 resolutions, WBC hotspots sharpen progressively from $1^\circ$ to $1/4^\circ$, with the resolution dependence most visible in the intermediate layer.
}
\label{fig:variance_ke}

\end{figure}

Figure~\ref{fig:overview_v1_v2}(e) provides direct visual evidence at 3100\,m: OM4 shows a nearly featureless temperature anomaly field, whereas Samudra exhibits pronounced zonal banding that closely mirrors the velocity structure shown in the panel below. Samudra 2 substantially suppresses this artifact, yielding a cleaner temperature field and improving correlation with OM4 from 0.27 to 0.37. This pattern is also evident in the variance maps. At 2000--7000\,m, Samudra's temperature variance map (Figure~\ref{fig:variance_thetao}) shows clear banding, while Samudra 2 largely removes these artifacts. The contrast is even stronger for salinity (Figure~\ref{fig:variance_so}): Samudra produces widespread spurious deep-ocean variance, whereas Samudra 2 suppresses most of it, though some residual variance persists in regions away from the WBC and Antarctic Circumpolar Current (ACC) hotspots, where physical variability alone does not account for it. By contrast, deep eddy kinetic energy (EKE) variance (Figure~\ref{fig:variance_ke}) shows neither banding nor widespread spurious variance, even in the original Samudra. This supports the interpretation of imprinting as a cross-variable artifact: because EKE is derived from the velocity fields themselves, its errors appear mainly as amplitude biases rather than alien spatial patterns projected from another variable. The suppression of imprinting by Samudra 2 also explains the smoother global mean trajectories in Figure~\ref{fig:overview_v1_v2}(b), where the high-frequency fluctuations present in Samudra are largely eliminated.

\begin{figure}[t]
\centering
\includegraphics[width=0.9\textwidth]{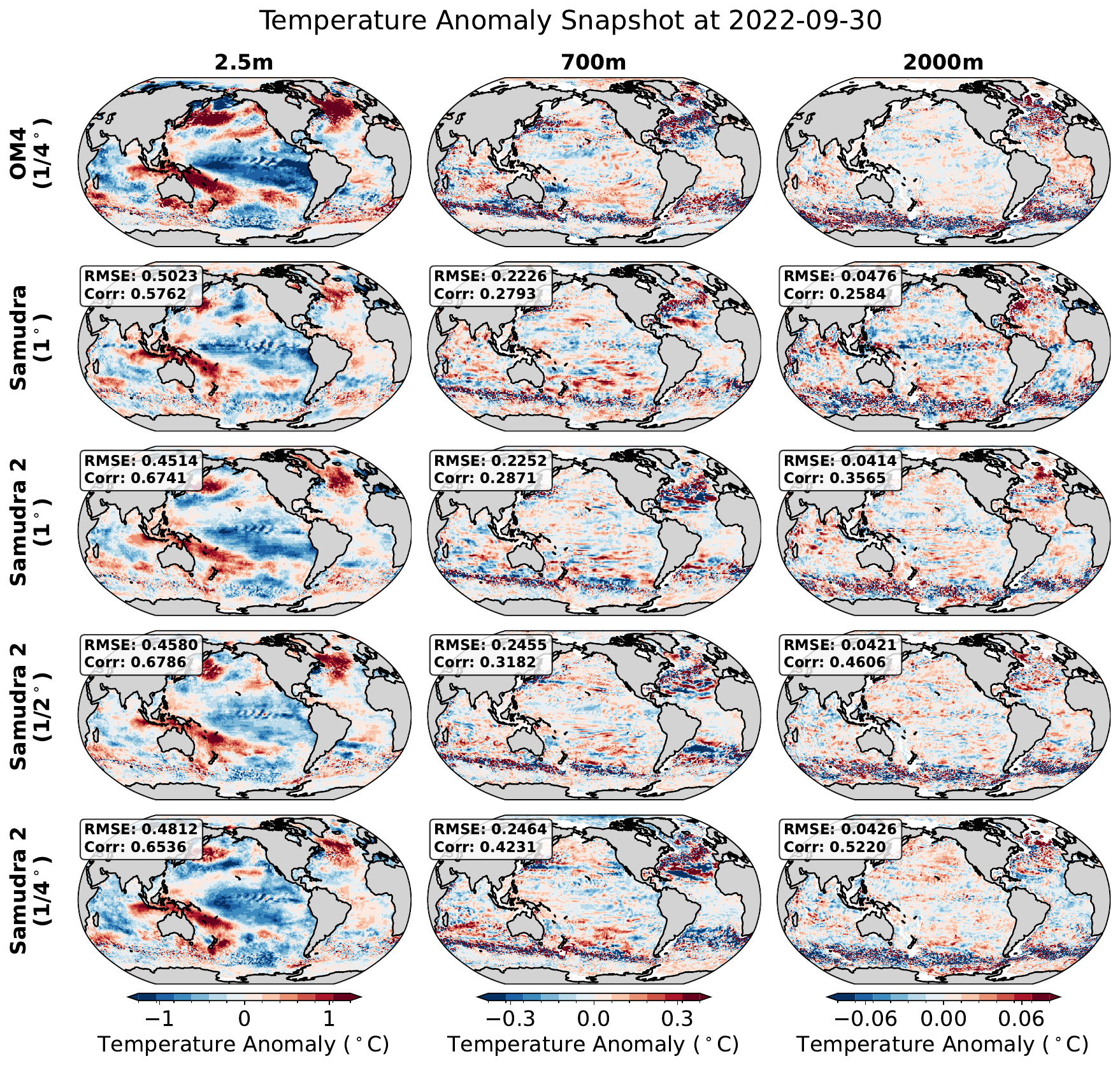}
\caption{Deseasonalized temperature anomaly snapshot at 2022-09-30, near the end of the 8-year rollout, at three depths (2.5\,m, 700\,m, 2000\,m) for OM4 ($1/4^\circ$), Samudra ($1^\circ$), and Samudra 2 at $1^\circ$, $1/2^\circ$, and $1/4^\circ$. RMSE and correlation values (against the OM4 truth at the corresponding resolution) are annotated for each panel.}
\label{fig:snapshot_thetao}

\end{figure}

\begin{figure}[t!]
\centering
\includegraphics[width=1\textwidth]{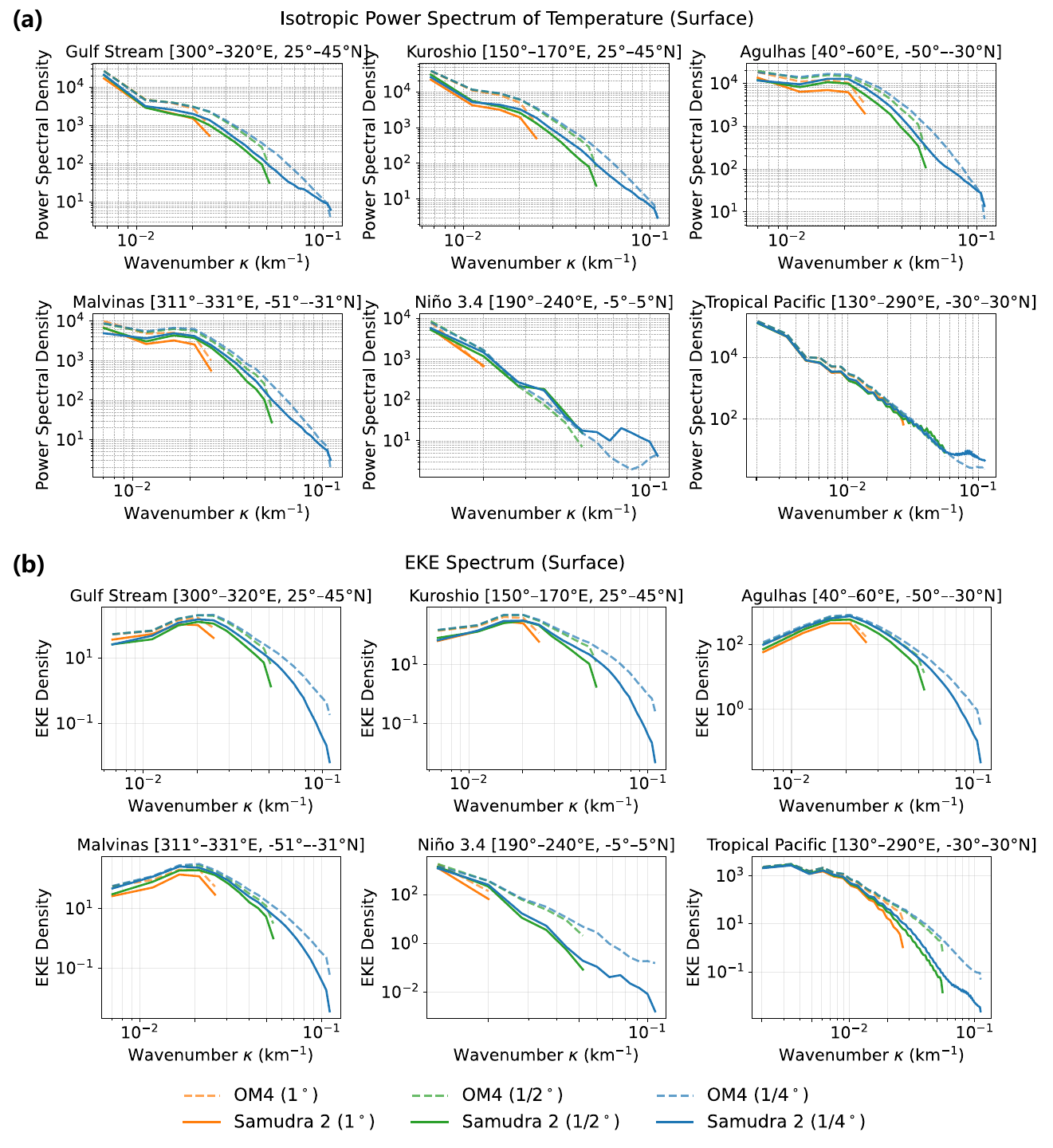}
\caption{Isotropic power spectra for six ocean regions (Gulf Stream, Kuroshio, Agulhas, Malvinas, Ni\~no 3.4, Tropical Pacific). Dashed lines show OM4 and solid lines Samudra 2 at $1^\circ$, $1/2^\circ$, and $1/4^\circ$.
(a) Surface temperature anomalies; higher resolutions extend the wavenumber range.
(b) EKE; similar to (a), but with a systematic small-scale energy deficit.}
\label{fig:spectrum_combined}

\end{figure}

\subsection{Multi-Resolution Evaluation}
\label{sec:variance}

We now scale Samudra 2 to $1/2^\circ$ and $1/4^\circ$ resolutions and examine whether increasing resolution yields higher fidelity, assessed through three complementary lenses: spatial variance structure, global-mean temporal tracking, and deseasonalized snapshot fields.
Across diagnostics, fidelity improves with resolution, with sharper variance fields, better temporal agreement, and more realistic deseasonalized snapshots.

Across the three Samudra 2 resolutions (Figure~\ref{fig:variance_thetao}; Figure~\ref{fig:variance_so}), variance fields resolve progressively finer spatial scales.
At $1^\circ$, the model already captures the large-scale hotspots of variance in the WBC region, though the fields remain relatively smooth.
At $1/2^\circ$ and $1/4^\circ$ resolutions, the WBC regions display progressively finer-grained variance structure, with the Gulf Stream, Kuroshio, and Agulhas regions all showing more spatially localized features.
The same trend holds for EKE (Figure~\ref{fig:variance_ke}), with the strongest resolution dependence: at $1^\circ$, WBC variance hotspots appear as broad, diffuse features, whereas at $1/4^\circ$ they become compact, spatially localized structures that closely match the OM4 reference, particularly in the intermediate layer, where the Gulf Stream, Kuroshio, and Agulhas regions all show markedly finer-scale variance.
This sharpening is more visually pronounced for EKE than for temperature or salinity, likely because kinetic energy is more spatially concentrated along fronts and eddy-rich corridors, making resolution-dependent gains easier to discern.
Regional temporal-variance metrics quantitatively confirm this trend: temperature variance correlation (Var Corr) improves across all depth ranges (e.g., global upper Var Corr: 0.83 at $1^\circ$ to 0.87 at $1/4^\circ$; intermediate: 0.54 to 0.78; deep: 0.68 to 0.83), with the largest relative gains at intermediate and deep levels.
We note that, because each resolution model is compared against its own OM4 truth, the higher-variance signal present in finer-resolution OM4 data may partly contribute to higher correlation scores;  the cross-resolution improvements should therefore be interpreted as reflecting both emulator capability and the higher-variance signal available at higher resolution.
KE shows a similar pattern, with the largest gains in the deep ocean (0.81 at $1^\circ$ to 0.88 at $1/4^\circ$).
Salinity remains the most challenging variable: intermediate and deep correlations improve with resolution but stay low, and isolated regions (e.g., Ni\~no~3.4 intermediate) exhibit negative Var Corr at all three resolutions, indicating that deep-ocean salinity variability is not yet reliably captured.

The spatial variance improvements are complemented by improved temporal fidelity of the global mean state.
Detrended global mean temperature time series across resolutions (Appendix 
Figure~\ref{fig:timeseries_thetao}) show that all three Samudra 2 models produce smooth trajectories with the high-frequency imprinting artifacts of the original Samudra substantially reduced, confirming that imprinting suppression generalizes across resolutions.
All three resolutions track the upper-ocean variability well ($R^2 = 0.84$--$0.92$). Below the upper ocean, performance remains limited but improves with resolution.
At intermediate depths, the $1/4^\circ$ model achieves $R^2 \approx 0.01$ (compared to $-1.60$ at $1^\circ$), approaching, but not yet reaching, useful predictive skill; an $R^2$ near zero means the model performs comparably to the temporal mean, which is a necessary threshold before deep-ocean predictions can be considered reliable.
At deep levels (2000--7000\,m), $R^2$ remains negative at all resolutions ($-16.14$ at $1^\circ$, $-9.98$ at $1/4^\circ$), confirming that deep-ocean temperature tracking is not yet solved despite the improvements from higher resolution and the dynamic loss.

The variance maps and global mean time series above summarize temporal and spatial variability respectively; we complement them with a deseasonalized temperature anomaly snapshot near the end of the rollout (Figure~\ref{fig:snapshot_thetao}), which shows the instantaneous spatial structure at the point of maximum autoregressive accumulation. The improvement with resolution is especially clear at depth, where higher-resolution models reproduce the anomaly field both more accurately overall and at progressively finer spatial scales; correlation at 2000\,m increases from 0.36 ($1^\circ$) to 0.52 ($1/4^\circ$), and the gain is more pronounced
in correlation than in RMSE. The same trends hold for velocity and EKE (Appendix~\ref{sec:multireso_results}, Figures~\ref{fig:snapshot_uo}--\ref{fig:snapshot_ke}): u-velocity correlation at 2000\,m improves from 0.39 ($1^\circ$) to 0.53 ($1/4^\circ$), and EKE correlation at 700\,m reaches 0.80 at $1/4^\circ$ compared to 0.64 at $1^\circ$.

Taken together, these results show that moving to higher resolution yields consistent fidelity gains, with the clearest improvements emerging below the upper ocean, where coarse-resolution models capture variability less effectively.

\subsection{Spectral Fidelity Scales with Resolution}
\label{sec:spectral_results}

We complement the physical-space diagnostics with spectral analysis, computing isotropic (spatial) and temporal power spectra of temperature and kinetic energy in six ocean regions (four WBC and two tropical regions).

Figure~\ref{fig:spectrum_combined}(a) shows isotropic power spectra of surface temperature anomalies.
At large scales, all Samudra 2 models track the OM4 truth closely; as resolution increases, the resolved wavenumber range extends substantially: the $1/4^\circ$ model captures spectral energy out to roughly four times the wavenumber of the $1^\circ$ model, accessing mesoscale dynamics entirely absent at coarser grids.
At each resolution, the emulator exhibits a systematic energy deficit that increases with wavenumber, a common characteristic of MSE-trained neural emulators~\citep{garg2026recipe}.
The isotropic EKE spectra (Figure~\ref{fig:spectrum_combined}b) show the same pattern, confirming that the spectral behavior generalizes across variables.
Temporal spectra and autocorrelation functions (Appendix~\ref{sec:spectral}, Figures~\ref{fig:spectrum_ke_temporal}--\ref{fig:acf_ssh}) further show that the spectral shape and decorrelation timescales of the OM4 truth are preserved at all resolutions.

\begin{figure}[t!]
\centering
\includegraphics[width=\textwidth]{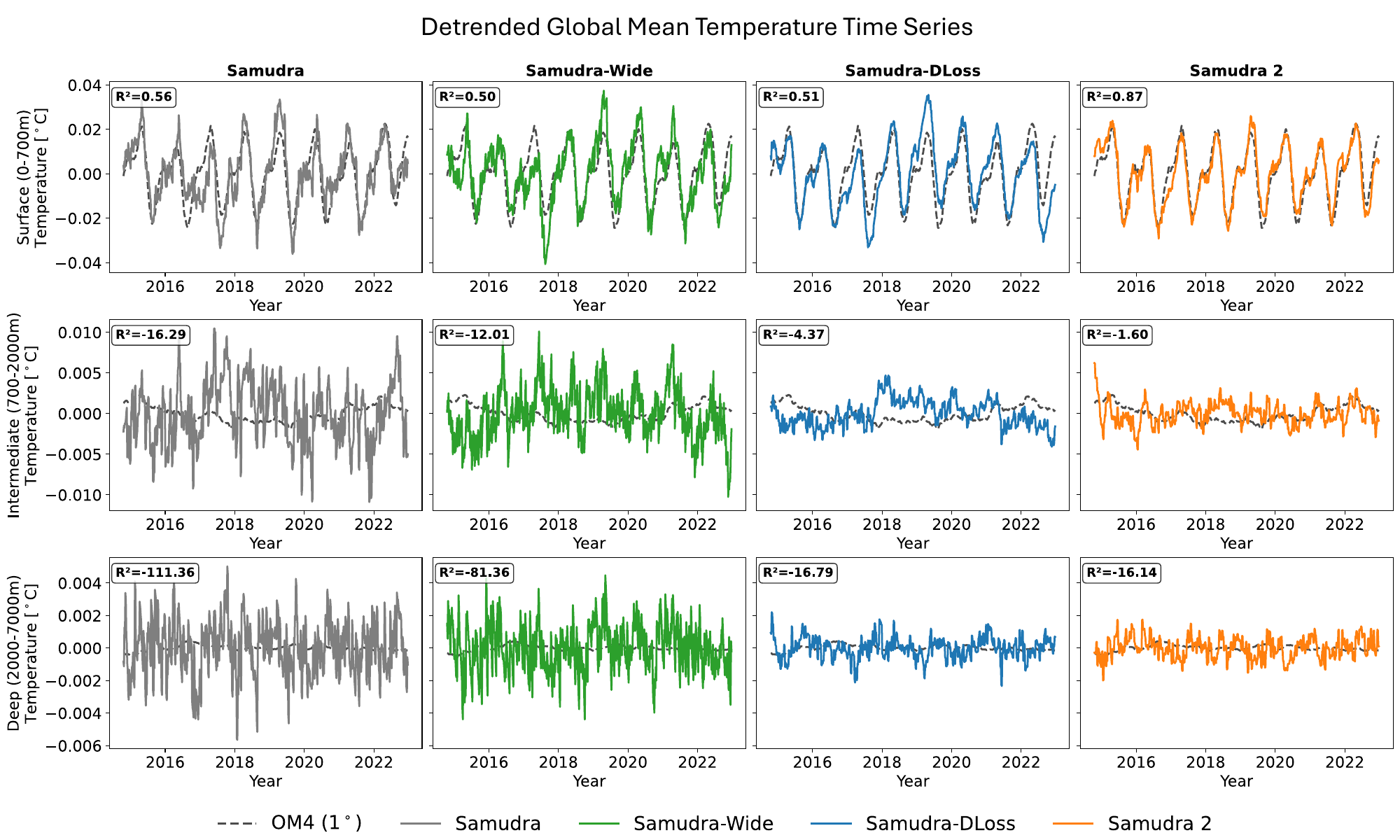}
\caption{Detrended global mean temperature time series for three depth ranges (Upper Ocean 0--700\,m, Intermediate Ocean 700--2000\,m, Deep Ocean 2000--7000\,m) for the four ablation variants. Dashed line: OM4 (truth) coarsened to $1^\circ$; grey: Samudra (baseline); green: Samudra-Wide; blue: Samudra-DLoss; orange: Samudra 2-$1^\circ$. $R^2$ values are annotated for each model at each depth.}
\label{fig:timeseries_thetao_only}

\end{figure}

\subsection{Ablation Study}
\label{sec:ablation}

Having shown that Samudra 2 improves fidelity relative to the original Samudra, we now study the contribution of the two key modifications to these gains. We compare four $1^\circ$ variants using detrended global mean time series and regional temporal-variance metrics: (1) \textbf{Samudra} (baseline; original architecture with standard MSE), (2) \textbf{Samudra-Wide} (wider ConvNeXt U-Net with standard MSE), (3) \textbf{Samudra-DLoss} (original architecture with dynamic loss), and (4) \textbf{Samudra 2} (wider architecture with dynamic loss).


\begin{table*}[t]
\centering
\footnotesize
\setlength{\tabcolsep}{3pt} 
\caption{Temporal-variance evaluation for potential temperature. All metrics are computed after removing the seasonal cycle from predictions and reference. \textbf{Var Corr} and \textbf{Var RMSE} are the area-weighted Pearson correlation and RMSE between the temporal-variance maps of prediction and reference. \textbf{Direct RMSE} and \textbf{Detrend RMSE} are the area-weighted means of
per-gridpoint temporal RMSE, computed on raw and linearly detrended fields respectively. Values are reported with three significant digits; the best in each row and depth range is highlighted in gray.}
\label{tab:temporal_variance_thetao}
\resizebox{\textwidth}{!}{%
\begin{tabular}{llcccccccccccc}
\toprule
\multirow{2}{*}{Region} & \multirow{2}{*}{Metric} & \multicolumn{4}{c}{Upper (0-700m)} & \multicolumn{4}{c}{Intermediate (700-2000m)} & \multicolumn{4}{c}{Deep (2000-7000m)} \\
\cmidrule(lr){3-6} \cmidrule(lr){7-10} \cmidrule(lr){11-14}
 &  & Samudra & Samudra-Wide & Samudra-DLoss & Samudra 2 & Samudra & Samudra-Wide & Samudra-DLoss & Samudra 2 & Samudra & Samudra-Wide & Samudra-DLoss & Samudra 2 \\
\midrule
Global & Var Corr & 0.797 & 0.796 & 0.788 & \best{0.826} & 0.677 & \best{0.690} & 0.665 & 0.538 & 0.656 & 0.646 & \best{0.704} & 0.681 \\
& Var RMSE & 0.160 & 0.165 & 0.158 & \best{0.153} & 0.0160 & \best{0.0158} & 0.0164 & 0.0233 & 0.00130 & 0.00131 & \best{0.00129} & 0.00134 \\
& Detrend RMSE & 0.189 & 0.182 & \best{0.174} & 0.176 & 0.0471 & 0.0460 & \best{0.0416} & 0.0416 & 0.0169 & 0.0174 & 0.0119 & \best{0.0119} \\
& Direct RMSE & 0.205 & 0.200 & \best{0.193} & 0.193 & 0.0503 & 0.0495 & 0.0460 & \best{0.0456} & 0.0176 & 0.0182 & 0.0130 & \best{0.0128} \\
\midrule
Gulf Stream & Var Corr & 0.891 & 0.862 & \best{0.915} & 0.844 & \best{0.573} & 0.497 & 0.239 & -0.0669 & \best{0.808} & 0.786 & 0.712 & 0.713 \\
& Var RMSE & 0.412 & 0.429 & \best{0.341} & 0.361 & \best{0.0212} & 0.0221 & 0.0665 & 0.121 & 5.18e-4 & \best{5.15e-4} & 6.67e-4 & 6.8e-4 \\
& Detrend RMSE & 0.391 & \best{0.382} & 0.399 & 0.437 & \best{0.140} & 0.141 & 0.176 & 0.201 & 0.0277 & 0.0279 & \best{0.0230} & 0.0237 \\
& Direct RMSE & \best{0.423} & 0.424 & 0.424 & 0.509 & \best{0.152} & 0.158 & 0.203 & 0.239 & 0.0309 & 0.0300 & 0.0272 & \best{0.0265} \\
\midrule
Kuroshio & Var Corr & 0.795 & 0.789 & \best{0.813} & 0.803 & 0.767 & 0.712 & \best{0.775} & 0.627 & 0.575 & 0.507 & \best{0.626} & 0.498 \\
& Var RMSE & 0.262 & 0.292 & 0.266 & \best{0.230} & \best{0.00499} & 0.00603 & 0.00544 & 0.00598 & 2.59e-4 & 2.99e-4 & \best{6.75e-5} & 1.41e-4 \\
& Detrend RMSE & 0.470 & 0.449 & \best{0.429} & 0.460 & 0.0719 & 0.0715 & \best{0.0673} & 0.0679 & 0.0148 & 0.0161 & \best{0.00880} & 0.0102 \\
& Direct RMSE & 0.519 & 0.479 & \best{0.474} & 0.500 & 0.0765 & 0.0757 & 0.0725 & \best{0.0717} & 0.0154 & 0.0166 & \best{0.00976} & 0.0111 \\
\midrule
Agulhas & Var Corr & 0.965 & 0.975 & \best{0.977} & 0.973 & \best{0.961} & 0.960 & 0.952 & 0.941 & 0.729 & 0.691 & \best{0.793} & 0.773 \\
& Var RMSE & 0.173 & 0.170 & \best{0.135} & 0.150 & 0.0161 & \best{0.0147} & 0.0169 & 0.0161 & \best{0.00122} & 0.00130 & 0.00128 & 0.00136 \\
& Detrend RMSE & 0.333 & 0.303 & \best{0.298} & 0.306 & 0.112 & 0.107 & \best{0.0943} & 0.0992 & 0.0381 & 0.0401 & \best{0.0306} & 0.0308 \\
& Direct RMSE & 0.375 & 0.365 & \best{0.325} & 0.339 & 0.116 & 0.110 & \best{0.0988} & 0.103 & 0.0394 & 0.0411 & \best{0.0318} & 0.0325 \\
\midrule
Malvinas & Var Corr & 0.546 & \best{0.760} & 0.586 & 0.598 & 0.470 & \best{0.630} & 0.477 & 0.365 & \best{0.616} & 0.608 & 0.475 & 0.518 \\
& Var RMSE & 0.177 & \best{0.150} & 0.159 & 0.163 & 0.00646 & \best{0.00558} & 0.00642 & 0.00697 & 0.00244 & \best{0.00232} & 0.00261 & 0.00263 \\
& Detrend RMSE & 0.353 & \best{0.313} & 0.330 & 0.323 & 0.0874 & 0.0909 & 0.0852 & \best{0.0787} & 0.0427 & 0.0425 & \best{0.0376} & 0.0383 \\
& Direct RMSE & 0.374 & \best{0.347} & 0.354 & 0.368 & 0.0952 & 0.0984 & 0.0897 & \best{0.0856} & 0.0456 & 0.0461 & 0.0432 & \best{0.0423} \\
\midrule
Niño 3.4 & Var Corr & 0.958 & 0.920 & 0.945 & \best{0.964} & 0.225 & 0.0151 & \best{0.477} & 0.451 & \best{0.146} & 0.133 & 0.136 & 0.101 \\
& Var RMSE & \best{0.0177} & 0.0282 & 0.0199 & 0.0305 & 0.00207 & 0.00239 & 5.66e-4 & \best{3.99e-4} & 6.51e-4 & 8.08e-4 & 2.4e-4 & \best{1.86e-4} \\
& Detrend RMSE & 0.176 & 0.177 & \best{0.157} & 0.157 & 0.0509 & 0.0542 & 0.0348 & \best{0.0339} & 0.0249 & 0.0267 & 0.0148 & \best{0.0135} \\
& Direct RMSE & 0.190 & 0.191 & \best{0.167} & 0.171 & 0.0513 & 0.0549 & 0.0353 & \best{0.0344} & 0.0251 & 0.0271 & 0.0151 & \best{0.0137} \\
\midrule
Tropical Pacific & Var Corr & 0.885 & 0.879 & 0.755 & \best{0.906} & 0.309 & 0.253 & \best{0.616} & 0.592 & 0.274 & 0.266 & 0.321 & \best{0.321} \\
& Var RMSE & \best{0.0665} & 0.0692 & 0.0841 & 0.0685 & 0.00729 & 0.00834 & 0.00625 & \best{0.00620} & 6.03e-4 & 7.03e-4 & 2.33e-4 & \best{2.15e-4} \\
& Detrend RMSE & 0.195 & 0.189 & 0.185 & \best{0.181} & 0.0430 & 0.0433 & 0.0374 & \best{0.0360} & 0.0167 & 0.0179 & 0.0107 & \best{0.0102} \\
& Direct RMSE & 0.214 & 0.210 & 0.208 & \best{0.199} & 0.0446 & 0.0454 & 0.0399 & \best{0.0379} & 0.0171 & 0.0184 & 0.0114 & \best{0.0106} \\
\bottomrule
\end{tabular}%
}
\end{table*}

Figure~\ref{fig:timeseries_thetao_only} shows detrended global mean
temperature time series for the four ablation variants. In the
upper ocean, neither modification alone improves over the baseline
($R^2 = 0.56$): Samudra-Wide ($R^2 = 0.50$) and Samudra-DLoss
($R^2 = 0.51$) both slightly underperform, yet their combination
in Samudra 2 raises $R^2$ to 0.87, indicating a strongly synergistic
interaction. At intermediate and deep levels, the dynamic loss is
the primary driver of improvement for temperature, reducing
intermediate-depth $R^2$ from $-16.29$ to $-4.37$ and deep $R^2$
from $-111.36$ to $-16.79$.
While these represent substantial relative gains, the values remain negative at all depths below 700\,m for all ablation variants, indicating that no current configuration produces deep-ocean predictions that outperform the temporal mean.
Salinity shows a similar pattern: the dynamic loss reduces deep-ocean $R^2$ magnitude from $-1423$ to $-72$, but these values underscore that deep salinity variability, whose anomalies are even smaller than those of temperature, is far from being reliably captured.
The wider architecture alone degrades intermediate-depth salinity ($R^2 = -99.46$), suggesting that increased capacity without proper loss balancing amplifies spurious deep-ocean gradients.

The regional breakdown (Table~\ref{tab:temporal_variance_thetao}; Appendix Tables~\ref{tab:temporal_variance_so} and~\ref{tab:temporal_variance_ke}) corroborates these findings: Samudra 2 achieves the best or near-best RMSE in most regions, though individual modifications sometimes yield higher Var Corr at intermediate and deep levels, indicating that the full combination optimizes overall error magnitude rather than uniformly improving every diagnostic.

In summary, the dynamic loss is the dominant driver of deep-ocean fidelity; the wider architecture alone does not yield clear improvements, but it provides the additional capacity necessary for a strongly synergistic interaction with the dynamic loss. In isolation, neither modification surpasses the baseline in the upper ocean, yet their combination in Samudra 2 yields $R^2 = 0.87$ (vs.\ $0.50$--$0.56$ for individual variants).
Samudra 2 does not uniformly perform across all metrics; individual modifications sometimes achieve higher correlation with variance or better deep-ocean salinity. Overall, however, the combination provides the most balanced performance across variables and depths.

\section{Discussion and Conclusion}
\label{sec:conclusion}

In this work, we present Samudra 2, which improves the fidelity of the Samudra ocean emulator with a wider ConvNeXt U-Net and a dynamic, inverse-error weighted loss function. Samudra 2 substantially reduces temporal variance collapse and imprinting artifacts, while improving key diagnostics, including Ni\~no 3.4, upper-ocean detrended temperature, deep-ocean temperature error (reduced approximately sevenfold), and pattern correlation at all depths.
In addition, the same architecture, trained independently at each of three
horizontal resolutions ($1^\circ$, $1/2^\circ$, and $1/4^\circ$) on GFDL OM4 simulation data, supports multi-year autoregressive ocean emulation. Higher-resolution models resolve progressively finer mesoscale features, producing sharper western boundary currents and improved eddy kinetic energy distributions.

Utilizing long rollouts on a single GPU, Samudra 2 makes it practical to conduct thousand-year long simulations to explore the ocean circulation's role in climate. It may make it possible to scale ensemble sizes for more robust uncertainty quantification for projections of regional sea-level, ocean heat uptake, and climate variability modes such as ENSO.  The emulator performs better in the upper ocean than in the deep ocean, consistent with the higher variability of the upper ocean. The emulator is therefore most readily applied to upper-ocean and surface-driven diagnostics, while deep-ocean fidelity remains an open challenge for which the underlying physics-based simulation remains the more reliable reference.

Several limitations point to directions for future work.
Deep-ocean predictions (below $\sim$700\,m) remain an open challenge, with negative $R^2$ for both temperature and salinity at most resolutions, owing to the extremely weak training signal at depth.
Promising remedies include spectral or adversarial loss functions that amplify the deep-ocean gradient signal and mitigate the systematic high-wavenumber energy deficit inherent to MSE minimization \citep{garg2026recipe, chattopadhyay2023challenges}, as recently demonstrated by atmospheric emulators using spectral-domain probabilistic losses \citep{bonev2025fourcastnet3}; hybrid physics-ML architectures that embed conservation laws or equation-of-state constraints \citep{kochkov2024neuralgcm}; and separate decoder heads for dynamically distinct variable groups to reduce cross-variable imprinting.
Beyond deep-ocean fidelity, cross-resolution transfer learning could reduce the computational cost of high-resolution training, and extending the emulator to additional variables (e.g., biogeochemistry, sea ice \citep{gregory2026advancing}), centennial timescales, and tighter atmospheric coupling \citep{duncan2025samudrace} would further broaden the applicability of neural ocean emulation.

More broadly, several bottlenecks confronted by Samudra 2 are shared with a wider class of ML and ML-for-physics problems: variance collapse over long autoregressive rollouts \citep{mathieu2016deep, hafner2023dreamerv3, lam2023graphcast, kochkov2024neuralgcm}, the high-wavenumber energy deficit of MSE-trained emulators reflecting neural networks' spectral bias \citep{rahaman2019spectral}, and the cross-variable loss imbalance addressed by adaptive loss balancing in multi-task learning \citep{kendall2018multi, chen2018gradnorm} and physics-informed neural networks \citep{wang2021understanding}.
Per-variable reweighting, first introduced for atmospheric forecasting \citep{keisler2022forecasting}, may generalize to other multi-variable autoregressive problems with order-of-magnitude variance imbalance.
We hope progress on these shared bottlenecks will accelerate neural climate emulation.

\acks{This project is supported by Schmidt Sciences, as part of the M$^2$LInES project. We also acknowledge support from the NSF CAIG program via grant 2530958. We thank NVIDIA for a GPU hardware grant, ongoing support, and helpful advice; Lambda (\url{https://lambda.ai/}) for a grant that provided the hardware for developing these models; and AWS for infrastructure grants, which provided data storage and engineering lifecycle support. This research was also supported in part through the NYU IT High Performance Computing resources, services, and staff expertise.}

\vskip 0.2in
\bibliography{sample}

@article{ravuri2021skilful,
  author  = {Suman Ravuri and Karel Lenc and others},
  title   = {Skilful precipitation nowcasting using deep generative models of radar},
  journal = {Nature},
  volume  = {597},
  number  = {7878},
  pages   = {672--677},
  year    = {2021},
  doi     = {10.1038/s41586-021-03854-z},
}

@Article{dheeshjith2025samudra,
  author  = {Dheeshjith, Surya and Subel, Adam and Adcroft, Alistair and Busecke, Julius and Fernandez-Granda, Carlos and Gupta, Shubham and Zanna, Laure},
  title   = {Samudra: An {AI} Global Ocean Emulator for Climate},
  journal = {Geophysical Research Letters},
  volume  = {52},
  number  = {10},
  pages   = {e2024GL114318},
  year    = {2025},
  doi     = {10.1029/2024GL114318},
}

@article{adcroft2019gfdl,
  author  = {Alistair Adcroft and Whit Anderson and others},
  title   = {The {GFDL} Global Ocean and Sea Ice Model {OM4.0}: Model Description and Simulation Characteristics},
  journal = {Journal of Advances in Modeling Earth Systems},
  volume  = {11},
  number  = {10},
  pages   = {3167--3211},
  year    = {2019},
}

@article{xu2023enhanced,
  title={Enhanced upper ocean warming projected by the eddy-resolving community earth system model},
  author={Xu, Gaopeng and Chang, Ping and others},
  journal={Geophysical Research Letters},
  volume={50},
  number={21},
  pages={e2023GL106100},
  year={2023},
  publisher={Wiley Online Library}
}

@article{pathak2022fourcastnet,
  author  = {Jaideep Pathak and Shashank Subramanian and others},
  title   = {{FourCastNet}: A Global Data-driven High-resolution Weather Forecasting Model},
  journal = {arXiv preprint arXiv:2202.11214},
  year    = {2022},
}

@article{dunne2024evolving,
  title={An evolving Coupled Model Intercomparison Project phase 7 (CMIP7) and Fast Track in support of future climate assessment},
  author={Dunne, John Patrick and others},
  journal={EGUsphere},
  volume={2024},
  pages={1--51},
  year={2024},
  publisher={Copernicus Publications G{\"o}ttingen, Germany}
}

@Article{bi2023pangu,
  author  = {Bi, Kaifeng and Xie, Lingxi and Zhang, Hengheng and Chen, Xin and Gu, Xiaotao and Tian, Qi},
  title   = {Accurate Medium-range Global Weather Forecasting with {3D} Neural Networks},
  journal = {Nature},
  volume  = {619},
  pages   = {533--538},
  year    = {2023},
}

@article{lam2023graphcast,
  author  = {Remi Lam and Alvaro Sanchez-Gonzalez and others},
  title   = {Learning Skillful Medium-range Global Weather Forecasting},
  journal = {Science},
  volume  = {382},
  number  = {6677},
  pages   = {1416--1421},
  year    = {2023},
}

@article{zhou2023self,
  title={A self-attention--based neural network for three-dimensional multivariate modeling and its skillful ENSO predictions},
  author={Zhou, Lu and Zhang, Rong-Hua},
  journal={Science Advances},
  volume={9},
  number={10},
  pages={eadf2827},
  year={2023},
  publisher={American Association for the Advancement of Science}
}

@Article{kochkov2024neuralgcm,
  author  = {Kochkov, Dmitrii and Yuval, Jann and others.},
  title   = {Neural General Circulation Models for Weather and Climate},
  journal = {Nature},
  volume  = {632},
  pages   = {1060--1066},
  year    = {2024},
}

@Article{ham2019deep,
  author  = {Ham, Yoo-Geun and Kim, Jeong-Hwan and Luo, Jing-Jia},
  title   = {Deep Learning for Multi-year {ENSO} Forecasts},
  journal = {Nature},
  volume  = {573},
  pages   = {568--572},
  year    = {2019},
}

@InProceedings{liu2022convnext,
  author    = {Liu, Zhuang and Mao, Hanzi and Wu, Chao-Yuan and Feichtenhofer, Christoph and Darrell, Trevor and Xie, Saining},
  title     = {A {ConvNet} for the 2020s},
  booktitle = {Proceedings of the IEEE/CVF Conference on Computer Vision and Pattern Recognition (CVPR)},
  pages     = {11976--11986},
  year      = {2022},
}

@InProceedings{ronneberger2015unet,
  author    = {Ronneberger, Olaf and Fischer, Philipp and Brox, Thomas},
  title     = {U-Net: Convolutional Networks for Biomedical Image Segmentation},
  booktitle = {Medical Image Computing and Computer-Assisted Intervention (MICCAI)},
  pages     = {234--241},
  year      = {2015},
}

@Article{mathieu2016deep,
  author  = {Mathieu, Michael and Couprie, Camille and LeCun, Yann},
  title   = {Deep Multi-scale Video Prediction Beyond Mean Square Error},
  journal = {arXiv preprint arXiv:1511.05440},
  year    = {2016},
}

@article{rasp2024weatherbench,
  author  = {Stephan Rasp and Stephan Hoyer and others},
  title   = {{WeatherBench} 2: A Benchmark for the Next Generation of Data-driven Global Weather Models},
  journal = {Journal of Advances in Modeling Earth Systems},
  volume  = {16},
  number  = {6},
  year    = {2024},
}

@Article{eyring2016cmip6,
  author  = {Eyring, Veronika and Bony, Sandrine and others},
  title   = {Overview of the {Coupled Model Intercomparison Project Phase 6 (CMIP6)}: Experimental Design and Organization},
  journal = {Geoscientific Model Development},
  volume  = {9},
  number  = {5},
  pages   = {1937--1958},
  year    = {2016},
}

@article{hewitt2020resolving,
  author  = {Helene T. Hewitt and Malcolm Roberts and others},
  title   = {Resolving and Parameterising the Ocean Mesoscale in Earth System Models},
  journal = {Current Climate Change Reports},
  volume  = {6},
  pages   = {137--152},
  year    = {2020},
}

@article{foxkemper2019challenges,
  author = {Baylor Fox-Kemper and Alistair Adcroft and others},
  title = {Challenges and Prospects in Ocean Circulation Models},
  journal = {Frontiers in Marine Science},
  volume = {6},
  pages = {65},
  year = {2019},
  doi = {10.3389/fmars.2019.00065}
}

@article{guan2022stable,
  author = {Yifei Guan and Ashesh Chattopadhyay and Adam Subel and Pedram Hassanzadeh},
  title = {Stable a posteriori {LES} of {2D} turbulence using convolutional neural networks: Backscattering analysis and generalization to higher {Re} via transfer learning},
  journal = {Journal of Computational Physics},
  volume = {458},
  pages = {111090},
  year = {2022},
  doi = {10.1016/j.jcp.2022.111090}
}

@article{frezat2022posteriori,
  author = {Hugo Frezat and Julien Le Sommer and Ronan Fablet and Guillaume Balarac and Redouane Lguensat},
  title = {A Posteriori Learning for Quasi-Geostrophic Turbulence Parametrization},
  journal = {Journal of Advances in Modeling Earth Systems},
  volume = {14},
  number = {11},
  pages = {e2022MS003124},
  year = {2022},
  doi = {10.1029/2022MS003124}
}

@Article{cui2025wenhai,
  author  = {Cui, Yingzhe and Wu, Ruohan and others},
  title   = {Forecasting the Eddying Ocean with a Deep Neural Network},
  journal = {Nature Communications},
  volume  = {16},
  pages   = {2268},
  year    = {2025},
  doi     = {10.1038/s41467-025-57389-2},
}

@Article{guo2025orcadl,
  author  = {Guo, Zijie and Lyu, Pumeng and others},
  title   = {Data-Driven Global Ocean Modeling for Seasonal to Decadal Prediction},
  journal = {Science Advances},
  volume  = {11},
  number  = {33},
  pages   = {eadu2488},
  year    = {2025},
  doi     = {10.1126/sciadv.adu2488},
}

@inproceedings{ioffe2015batch,
  title={Batch normalization: Accelerating deep network training by reducing internal covariate shift},
  author={Ioffe, Sergey and Szegedy, Christian},
  booktitle={International conference on machine learning},
  pages={448--456},
  year={2015},
  organization={pmlr}
}

@Article{perezhogin2025generalizable,
  author  = {Perezhogin, Pavel and Adcroft, Alistair and Zanna, Laure},
  title   = {Generalizable Neural-Network Parameterization of Mesoscale Eddies in Idealized and Global Ocean Models},
  journal = {Geophysical Research Letters},
  volume  = {52},
  year    = {2025},
  doi     = {10.1029/2025GL117046},
}

@Article{perezhogin2024stable,
  author  = {Perezhogin, Pavel and Zhang, Cheng and Adcroft, Alistair and Fernandez-Granda, Carlos and Zanna, Laure},
  title   = {A Stable Implementation of a Data-Driven Scale-Aware Mesoscale Parameterization},
  journal = {Journal of Advances in Modeling Earth Systems},
  volume  = {16},
  number  = {10},
  pages   = {e2023MS004104},
  year    = {2024},
  doi     = {10.1029/2023MS004104},
}

@Article{guillaumin2021stochastic,
  author  = {Guillaumin, Arthur P. and Zanna, Laure},
  title   = {Stochastic-Deep Learning Parameterization of Ocean Momentum Forcing},
  journal = {Journal of Advances in Modeling Earth Systems},
  volume  = {13},
  number  = {9},
  pages   = {e2021MS002534},
  year    = {2021},
  doi     = {10.1029/2021MS002534},
}

@Article{duncan2025samudrace,
  author  = {Duncan, James P. C. and Wu, Elynn and others},
  title   = {{SamudrACE}: Fast and Accurate Coupled Climate Modeling with {3D} Ocean and Atmosphere Emulators},
  journal = {arXiv preprint arXiv:2509.12490},
  year    = {2025},
}

@Article{dheeshjith2024transfer,
  author  = {Dheeshjith, Surya and Subel, Adam and others.},
  title   = {Transfer Learning for Emulating Ocean Climate Variability across {CO\textsubscript{2}} Forcing},
  journal = {arXiv preprint arXiv:2405.18585},
  year    = {2024},
}

@Article{hallberg2013resolution,
  author  = {Hallberg, Robert},
  title   = {Using a Resolution Function to Regulate Parameterizations of Oceanic Mesoscale Eddy Effects},
  journal = {Ocean Modelling},
  volume  = {72},
  pages   = {92--103},
  year    = {2013},
  doi     = {10.1016/j.ocemod.2013.08.007},
}

@InProceedings{kingma2015adam,
  author    = {Kingma, Diederik P. and Ba, Jimmy},
  title     = {Adam: A Method for Stochastic Optimization},
  booktitle = {Proceedings of the 3rd International Conference on Learning Representations (ICLR)},
  year      = {2015},
}

@InProceedings{loshchilov2017sgdr,
  author    = {Loshchilov, Ilya and Hutter, Frank},
  title     = {{SGDR}: Stochastic Gradient Descent with Warm Restarts},
  booktitle = {Proceedings of the 5th International Conference on Learning Representations (ICLR)},
  year      = {2017},
}

@article{watt2025ace2,
  title={ACE2: accurately learning subseasonal to decadal atmospheric variability and forced responses},
  author={Watt-Meyer, Oliver and others},
  journal={npj Climate and Atmospheric Science},
  volume={8},
  number={1},
  pages={205},
  year={2025},
  publisher={Nature Publishing Group UK London}
}

@article{ulyanov2016instance,
  title={Instance normalization: The missing ingredient for fast stylization},
  author={Ulyanov, Dmitry and Vedaldi, Andrea and Lempitsky, Victor},
  journal={arXiv preprint arXiv:1607.08022},
  year={2016}
}

@Article{garg2026recipe,
  author  = {Garg, Piyush and Gergel, Diana R. and Shao, Andrew E. and Yacalis, Galen J.},
  title   = {The Recipe Matters More Than the Kitchen: Mathematical Foundations of the {AI} Weather Prediction Pipeline},
  journal = {arXiv preprint arXiv:2604.01215},
  year    = {2026},
}

@Article{polyak1992acceleration,
  author  = {Polyak, Boris T. and Juditsky, Anatoli B.},
  title   = {Acceleration of Stochastic Approximation by Averaging},
  journal = {SIAM Journal on Control and Optimization},
  volume  = {30},
  number  = {4},
  pages   = {838--855},
  year    = {1992},
}

@Article{hafner2023dreamerv3,
  author  = {Hafner, Danijar and Pasukonis, Jurgis and Ba, Jimmy and Lillicrap, Timothy},
  title   = {Mastering Diverse Domains through World Models},
  journal = {arXiv preprint arXiv:2301.04104},
  year    = {2023},
}

@Article{wattmeyer2023ace,
  author  = {Watt-Meyer, Oliver and Dresdner, Gideon and others},
  title   = {{ACE}: A fast, skillful learned global atmospheric model for climate prediction},
  journal = {arXiv preprint arXiv:2310.02074},
  year    = {2023},
}

@article{zanna2020data,
  author    = {Zanna, Laure and Bolton, Thomas},
  title     = {Data-Driven Equation Discovery of Ocean Mesoscale Closures},
  journal   = {Geophysical Research Letters},
  volume    = {47},
  number    = {17},
  pages     = {e2020GL088376},
  year      = {2020},
  doi       = {10.1029/2020GL088376}
}

@article{yan2024choice,
  author    = {Yan, F. E. and Mak, J. and Wang, Y.},
  title     = {On the Choice of Training Data for Machine Learning of
               Geostrophic Mesoscale Turbulence},
  journal   = {Journal of Advances in Modeling Earth Systems},
  volume    = {16},
  number    = {2},
  pages     = {e2023MS003915},
  year      = {2024},
  doi       = {10.1029/2023MS003915}
}

@article{maddison2026online,
  author    = {Maddison, James R.},
  title     = {Online Learning in Idealized Ocean Gyres},
  journal   = {Journal of Advances in Modeling Earth Systems},
  volume    = {18},
  number    = {2},
  pages     = {e2024MS004883},
  year      = {2026},
  doi       = {10.1029/2024MS004883}
}

@article{wang2024applications,
  author    = {Wang, Guosong and Hou, Min and others},
  title     = {Applications of Deep Learning Parameterization of Ocean Momentum Forcing},
  journal   = {arXiv preprint arXiv:2406.03659},
  year      = {2024},
  doi       = {10.48550/arXiv.2406.03659}
}

@article{zhu2022physics,
  author    = {Zhu, Yuchao and Zhang, Rong-Hua and Moum, James N. and
               Wang, Fan and Li, Xiaofeng and Li, Delei},
  title     = {Physics-Informed Deep-Learning Parameterization of Ocean Vertical Mixing
               Improves Climate Simulations},
  journal   = {National Science Review},
  volume    = {9},
  number    = {8},
  pages     = {nwac044},
  year      = {2022},
  doi       = {10.1093/nsr/nwac044}
}

@article{bodner2025data,
  author    = {Bodner, Abigail S. and Balwada, Dhruv and Zanna, Laure},
  title     = {A Data-Driven Approach for Parameterizing Ocean Submesoscale Buoyancy Fluxes},
  journal   = {Journal of Advances in Modeling Earth Systems},
  year      = {2025},
  note      = {Early View},
  doi       = {10.1029/2025MS004991}
}

@inproceedings{rahaman2019spectral,
  author    = {Rahaman, Nasim and Baratin, Aristide and Arpit, Devansh and
               others},
  title     = {On the Spectral Bias of Neural Networks},
  booktitle = {Proceedings of the 36th International Conference on Machine Learning (ICML)},
  pages     = {5301--5310},
  year      = {2019}
}

@inproceedings{kendall2018multi,
  author    = {Kendall, Alex and Gal, Yarin and Cipolla, Roberto},
  title     = {Multi-Task Learning Using Uncertainty to Weigh Losses for
               Scene Geometry and Semantics},
  booktitle = {Proceedings of the IEEE Conference on Computer Vision and
               Pattern Recognition (CVPR)},
  pages     = {7482--7491},
  year      = {2018}
}

@inproceedings{chen2018gradnorm,
  author    = {Chen, Zhao and Badrinarayanan, Vijay and Lee, Chen-Yu and
               Rabinovich, Andrew},
  title     = {{GradNorm}: Gradient Normalization for Adaptive Loss Balancing
               in Deep Multitask Networks},
  booktitle = {Proceedings of the 35th International Conference on Machine
               Learning (ICML)},
  pages     = {794--803},
  year      = {2018}
}

@article{wang2021understanding,
  author    = {Wang, Sifan and Teng, Yujun and Perdikaris, Paris},
  title     = {Understanding and Mitigating Gradient Flow Pathologies in
               Physics-Informed Neural Networks},
  journal   = {SIAM Journal on Scientific Computing},
  volume    = {43},
  number    = {5},
  pages     = {A3055--A3081},
  year      = {2021},
  doi       = {10.1137/20M1318043}
}

@article{chattopadhyay2023challenges,
  author    = {Chattopadhyay, Ashesh and Sun, Y. Qiang and Hassanzadeh, Pedram},
  title     = {Challenges of Learning Multi-Scale Dynamics with {AI} Weather Models:
               Implications for Stability and One Solution},
  journal   = {arXiv preprint arXiv:2304.07029},
  year      = {2023},
  doi       = {10.48550/arXiv.2304.07029}
}

@article{bonev2025fourcastnet3,
  author    = {Bonev, Boris and Kurth, Thorsten and others},
  title     = {{FourCastNet 3}: A Geometric Approach to Probabilistic Machine-Learning
               Weather Forecasting at Scale},
  journal   = {arXiv preprint arXiv:2507.12144},
  year      = {2025},
  doi       = {10.48550/arXiv.2507.12144}
}

@inproceedings{zagoruyko2016wide,
  author    = {Zagoruyko, Sergey and Komodakis, Nikos},
  title     = {Wide Residual Networks},
  booktitle = {Proceedings of the British Machine Vision Conference (BMVC)},
  year      = {2016},
  doi       = {10.5244/C.30.87}
}

@article{gregory2026advancing,
  title={Advancing global sea ice prediction capabilities using a fully coupled climate model with integrated machine learning},
  author={Gregory, William and Bushuk, Mitchell and Zhang, Yong-Fei and Adcroft, Alistair and Zanna, Laure and McHugh, Colleen and Jia, Liwei},
  journal={Science Advances},
  volume={12},
  number={1},
  pages={eady8957},
  year={2026},
  publisher={American Association for the Advancement of Science}
}

@article{keisler2022forecasting,
  title={Forecasting global weather with graph neural networks},
  author={Keisler, Ryan},
  journal={arXiv preprint arXiv:2202.07575},
  year={2022}
}

@article{siddiqui2024exploring,
  author    = {Siddiqui, Shoaib Ahmed and others},
  title     = {Exploring the Design Space of Deep-Learning-Based Weather Forecasting Systems},
  journal   = {arXiv preprint arXiv:2410.07472},
  year      = {2024},
  doi       = {10.48550/arXiv.2410.07472}
}

\clearpage
\appendix

\renewcommand{\topfraction}{0.9}
\renewcommand{\bottomfraction}{0.9}
\renewcommand{\textfraction}{0.05}
\renewcommand{\floatpagefraction}{0.8}
\setcounter{topnumber}{3}
\setcounter{bottomnumber}{3}
\setcounter{totalnumber}{6}

\section{Implementation Details}
\label{sec:impl_details}

This appendix summarizes implementation details, derived quantities, and evaluation metrics.
Unless otherwise noted, all quantities are computed on regular latitude-longitude grids after conservative regridding from the native OM4 tripolar output with \texttt{xesmf}.

\subsection{Training Setup and Computational Considerations}
All models are optimized with Adam \citep{kingma2015adam} using an initial learning rate of $6 \times 10^{-4}$, batch size 4, cosine annealing \citep{loshchilov2017sgdr} over 70 epochs (no warmup), and gradient clipping at 1.0.
No weight decay or dropout is applied.
The final model is the last-epoch checkpoint rather than the one with the lowest validation loss, since single-step validation loss is not fully predictive of long-horizon rollout performance; in practice, we observed no significant overfitting over the 70-epoch schedule.
An exponential moving average (EMA) of the parameters ($\beta = 0.999$, with ramp-up $\tilde{\beta}_n = \min(\beta,\, (1+n)/(10+n))$ \citep{polyak1992acceleration}) is maintained and used for all evaluation.
All prognostic and boundary variables are normalized to zero mean and unit variance using per-channel training-period statistics; land points are filled with zeros.

Moving from $1^\circ$ ($180 \times 360$) to $1/4^\circ$ ($720 \times 1440$) increases grid points by ${\sim}16\times$, affecting both data loading and peak GPU memory.
At larger data sizes, data loading (disk to host, then host to GPU) becomes a substantial fraction of training time; we improve this component over Samudra by loading multiple channels in parallel (both disk-to-CPU and CPU-to-GPU), reducing expensive indexing and reshaping, and moving normalization and masking to the GPU.
Peak memory, dominated by activations and gradients that scale with input size, is reduced 60\% via gradient checkpointing on selected layers at a modest 20\% increase in GPU time, allowing both $1^\circ$ and $1/2^\circ$ to fit within the 80\,GB per-GPU budget.
At $1/4^\circ$, we further reduce batch size from 4 to 1 per GPU to fit the $4\times$ larger data, replace batch norm~\citep{ioffe2015batch} with instance norm~\citep{ulyanov2016instance}, accumulate gradients over 4 forward passes before each parameter update, and switch from float32 to bfloat16 to further reduce memory.
These changes maintain training dynamics comparable to the $1^\circ$ and $1/2^\circ$ configurations.
Training uses PyTorch \texttt{DistributedDataParallel} across 8 A100 GPUs; Table~\ref{tab:training_cost} summarizes hardware and wall-clock costs.

\begin{table}[h]
\centering
\caption{Training cost per resolution (70 epochs each).}
\label{tab:training_cost}
\begin{tabular}{lccc}
\hline
Resolution & Grid size & GPUs & Approx.\ time per epoch \\
\hline
$1^\circ$     & $180 \times 360$   & 8$\times$ A100 (80\,GB) &  8 minutes \\
$1/2^\circ$   & $360 \times 720$   & 8$\times$ A100 (80\,GB) & 15 minutes \\
$1/4^\circ$   & $720 \times 1440$  & 8$\times$ A100 (80\,GB) & 50 minutes \\
\hline
\end{tabular}
\end{table}

\subsection{Grid and Vertical Coordinate}
\label{sec:grid_metrics}

The regular grid has $N_y \times N_x$ cells with centers $(\phi_i, \lambda_j)$ and boundary coordinates $(\phi^b, \lambda^b)$.
Grid spacings are $\Delta x_{i,j} = R_E \cos(\phi_i)\,\Delta\lambda^b_j$ and $\Delta y_{i,j} = R_E\,\Delta\phi^b_i$ ($R_E = 6.371 \times 10^6$\,m); cell areas $A_{i,j}$ are computed via \texttt{xesmf.util.cell\_area}.
The vertical coordinate uses 19 depth levels at
$z_\ell \in \{2.5,\, 10,\, 22.5,\, 40,\, 65,\, 105,\, 165,\, 250,\, 375,\, 550,\, 775,\, 1050,\, 1400,\, 1850,\, 2400,\, 3100,$ $4000,\, 5000,\, 6000\}\;\text{m}$,
with layer interfaces at
$z^b \in \{0,\, 5,\, 15,\, 30,\, 50,\, 80,\, 130,\, 200,\, 300,\, 450,$ $650,\, 900,\, 1200,\, 1600,\, 2100,\, 2700,\, 3500,\, 4500,\, 5500,\, 6750\}\;\text{m}$.
Layer thickness is $\Delta z_\ell = z^b_{\ell+1} - z^b_\ell$, and $w_{i,j,\ell} \in \{0,1\}$ is the wet mask.
Three standard depth slices are used throughout: Upper Ocean (0--700\,m), Intermediate Ocean (700--2000\,m), and Deep Ocean (2000--7000\,m).

\subsection{Derived Quantities}
\label{sec:derived_quantities}

\paragraph{Vertical averaging.}
\label{sec:vert_avg}
The depth-averaged value of a 3D field $\psi(t,i,j,\ell)$ over a depth range $[z_{\min}, z_{\max}]$ is:
\begin{equation}
    \bar{\psi}(t, i, j) = \frac{\sum_{\ell \in \mathcal{S}} \psi(t, i, j, \ell)\,\Delta z_\ell\,w_{i,j,\ell}}{\sum_{\ell \in \mathcal{S}} \Delta z_\ell\,w_{i,j,\ell}},
    \label{eq:vert_avg}
\end{equation}
where $\mathcal{S} = \{\ell : z_{\min} \le z_\ell \le z_{\max}\}$.

\paragraph{Ocean heat content (OHC).}
\label{sec:ohc_calc}
$\text{OHC}(t,i,j) = \rho_0 c_p \sum_{\ell \in \mathcal{S}} \theta(t,i,j,\ell)\,\Delta z_\ell\,w_{i,j,\ell}$,
where $\rho_0 = 1035$\,kg\,m$^{-3}$ and $c_p = 3850$\,J\,kg$^{-1}$\,K$^{-1}$, yielding units of J\,m$^{-2}$ (scaled by $10^{-21}$ for ZJ).

\paragraph{Kinetic energy (KE) and eddy kinetic energy (EKE).}
\label{sec:ke_calc}
\begin{align}
    \text{KE}(t,i,j,\ell) &= \tfrac{1}{2}(u^2 + v^2), \label{eq:ke}\\
    \text{EKE}(t,i,j,\ell) &= \tfrac{1}{2}(u'^2 + v'^2), \label{eq:eke}
\end{align}
where $u' = u - \overline{u}$ and $v' = v - \overline{v}$ are anomalies relative to the temporal mean.
Note that Figure~\ref{fig:variance_ke} shows the temporal variance of depth-averaged KE (Eq.~\ref{eq:temporal_var}), not time-averaged EKE.

\subsection{Evaluation Diagnostics}
\label{sec:eval_diagnostics}

\subsubsection{Global Mean Time Series}
\label{sec:global_mean_ts}

The volume-weighted global mean of a 3D field over a depth slice is:
\begin{equation}
    \langle\psi\rangle(t) = \frac{\sum_{i,j}\sum_{\ell \in \mathcal{S}} \psi(t,i,j,\ell)\,\Delta z_\ell\,A_{i,j}\,w_{i,j,\ell}}{\sum_{i,j}\sum_{\ell \in \mathcal{S}} \Delta z_\ell\,A_{i,j}\,w_{i,j,\ell}}.
    \label{eq:global_mean}
\end{equation}

\subsubsection{Linear Detrending}
\label{sec:detrending}

All detrending removes a least-squares linear fit.
For a 1D time series $y(t)$, the detrended series is $y_{\text{det}}(t) = y(t) - [\hat{\beta}(t - \bar{t}) + \bar{y}]$, where
\begin{equation}
    \hat{\beta} = \frac{\sum_t (t - \bar{t})(y(t) - \bar{y})}{\sum_t (t - \bar{t})^2}.
    \label{eq:slope_1d}
\end{equation}
For spatially resolved fields, the same procedure is applied independently at each grid point:
\begin{equation}
    \psi_{\text{det}}(t, i, j) = \psi(t, i, j) - [\hat{\beta}_{i,j}(t - \bar{t}) + \overline{\psi}_{i,j}].
    \label{eq:detrend_spatial}
\end{equation}
Before computing spatial power spectra, a 2D linear plane $(\hat{a}\,x_j + \hat{b}\,y_i + \hat{c})$ is removed from each snapshot, where $(x_j, y_i) \in [-1,1]$.

\subsubsection{Temporal Variance Maps}
\label{sec:var_maps}

The temporal variance of a depth-averaged field is:
\begin{equation}
    \sigma^2_\psi(i, j) = \frac{1}{T}\sum_{t=1}^{T} [\bar{\psi}(t, i, j) - \overline{\bar{\psi}}(i, j)]^2.
    \label{eq:temporal_var}
\end{equation}

\subsubsection{Zonal Mean Profiles}
\label{sec:zonal_mean_calc}

The zonal mean of a 3D field within basin $B$ is:
\begin{equation}
    \langle\psi\rangle_{\text{zonal}}(i, \ell) = \frac{\sum_{j} \psi(i, j, \ell)\,M^B_{i,j}\,w_{i,j,\ell}\,\Delta x_{i,j}}{\sum_{j} M^B_{i,j}\,w_{i,j,\ell}\,\Delta x_{i,j}},
    \label{eq:zonal_mean}
\end{equation}
where $M^B_{i,j} \in \{0,1\}$ is the basin mask.
For OHC zonal cross-sections, the per-level contribution $q(i,j,\ell) = \rho_0 c_p \theta(i,j,\ell) \Delta z_\ell$ is zonally averaged at each depth level without prior vertical summation.

\subsubsection{Ni\~no 3.4 Index}
\label{sec:nino34_calc}

The Ni\~no 3.4 index is computed from the surface temperature ($z = 2.5$\,m) over the box $\lambda \in [190^\circ\text{E}, 240^\circ\text{E}]$, $\phi \in [5^\circ\text{S}, 5^\circ\text{N}]$ by (1)~subtracting a pentad-of-year climatology to obtain anomalies $\theta'$, (2)~applying a 150-day running mean (30 five-day steps), and (3)~computing the area-weighted spatial average.
The first 30 time steps are discarded to avoid edge effects.

\subsection{Spectral Analysis}
\label{sec:spectrum_calc}

\paragraph{Isotropic power spectrum.}
For a 2D field on a rectangular subregion ($H \times W$, spacings $\Delta x$, $\Delta y$): remove the spatial mean and a linear plane, apply a 2D Hann window, compute the 2D real FFT with \texttt{forward} normalization, correct for windowing, then azimuthally average the PSD into $N_b = \lfloor\min(H,W)/4\rfloor$ radial wavenumber bins over $[0, k_{\text{Nyq}}]$.
The plotted quantity is $k_b \cdot \overline{\text{PSD}}(k_b)$ (variance-preserving form).
For regional spectra, the field is extracted over the region's box and time-averaged before applying this procedure.

The six spectrum regions are: Gulf Stream ($[300^\circ, 320^\circ] \times [25^\circ, 45^\circ]$), Kuroshio ($[150^\circ, 170^\circ] \times [25^\circ, 45^\circ]$), Agulhas ($[40^\circ, 60^\circ] \times [-50^\circ, -30^\circ]$), Malvinas ($[311^\circ, 331^\circ] \times [-51^\circ, -31^\circ]$), Ni\~no 3.4 ($[190^\circ, 240^\circ] \times [-5^\circ, 5^\circ]$), and Tropical Pacific ($[130^\circ, 290^\circ] \times [-30^\circ, 30^\circ]$).

\paragraph{EKE spectrum.}
\label{sec:eke_spectrum}
$S_{\text{KE}}(k_b) = \frac{1}{2}[S_u(k_b) + S_v(k_b)]$, where $S_u$ and $S_v$ are the isotropic spectra of each velocity component.

\paragraph{Temporal power spectrum.}
\label{sec:temporal_spectrum}
For each grid point in a region, compute $\text{PSD}(f; i,j) = |\text{rfft}(\psi(t,i,j))|^2 \cdot T_{\text{phys}}$ and spatially average.
The temporal EKE spectrum combines both velocity components as $\frac{1}{2}[\text{PSD}_u(f) + \text{PSD}_v(f)]$.

\subsection{Quantitative Evaluation Metrics}
\label{sec:quant_metrics}

All area-weighted metrics are computed on the 2D grid after depth-averaging (Eq.~\ref{eq:vert_avg}).

\paragraph{Area-weighted RMSE.}
\label{sec:rmse}
\begin{equation}
    \text{RMSE} = \sqrt{\frac{\sum_{i,j} [\psi^{\text{truth}}(i,j) - \psi^{\text{pred}}(i,j)]^2 A_{i,j}}{\sum_{i,j} A_{i,j}}}.
    \label{eq:rmse}
\end{equation}

\paragraph{Area-weighted Pearson correlation.}
\label{sec:correlation}
\begin{equation}
    r = \frac{\sum_{i,j}(\psi^{\text{truth}}_{i,j} - \langle\psi^{\text{truth}}\rangle_A)(\psi^{\text{pred}}_{i,j} - \langle\psi^{\text{pred}}\rangle_A)\,A_{i,j}}{\sqrt{\sum_{i,j}(\psi^{\text{truth}}_{i,j} - \langle\psi^{\text{truth}}\rangle_A)^2 A_{i,j}} \;\sqrt{\sum_{i,j}(\psi^{\text{pred}}_{i,j} - \langle\psi^{\text{pred}}\rangle_A)^2 A_{i,j}}},
    \label{eq:pearson}
\end{equation}
where $\langle\psi\rangle_A = \sum_{i,j}\psi_{i,j} A_{i,j} / \sum_{i,j} A_{i,j}$.

\paragraph{Coefficient of determination ($R^2$).}
\label{sec:r2}
$R^2 = 1 - \sum_t [y^{\text{truth}}(t) - y^{\text{pred}}(t)]^2 \,/\, \sum_t [y^{\text{truth}}(t) - \overline{y^{\text{truth}}}]^2$.
Note that $R^2$ can be negative when predictions are worse than the temporal mean, which commonly occurs for deep-ocean variables.

\paragraph{Temporal variance metrics.}
\label{sec:variance_metrics}
The temporal-variance tables report four metrics per variable, region, and depth slice:
(1)~\textbf{Var Corr}: area-weighted Pearson correlation between the truth and prediction temporal variance maps (Eq.~\ref{eq:temporal_var});
(2)~\textbf{Var RMSE}: area-weighted RMSE between the two variance maps;
(3)~\textbf{Direct RMSE}: area-weighted mean of per-gridpoint temporal RMSE;
(4)~\textbf{Detrend RMSE}: same as Direct RMSE but applied to per-gridpoint linearly detrended fields (Eq.~\ref{eq:detrend_spatial}).
Metrics are computed over seven regions: Global plus the six spectrum regions above.

\paragraph{Ni\~no 3.4 summary metrics.}
\label{sec:nino_metrics}
$R^2$, Pearson correlation, MAE, and RMSE between the truth and predicted Ni\~no 3.4 time series.
\section{Additional Results}
\label{sec:add_results}

This appendix provides supplementary diagnostics for the four models compared in the main text: the original Samudra baseline at $1^\circ$ resolution and Samudra 2 at $1^\circ$, $1/2^\circ$, and $1/4^\circ$ resolutions. Each model is evaluated against the OM4 truth at its corresponding resolution over the test period ($\sim$2014--2022, $\sim$580 five-day steps).

\subsection{Multi-Resolution Spatial Fields}
\label{sec:multireso_results}

\begin{figure}[htbp]
\centering
\includegraphics[width=\textwidth]{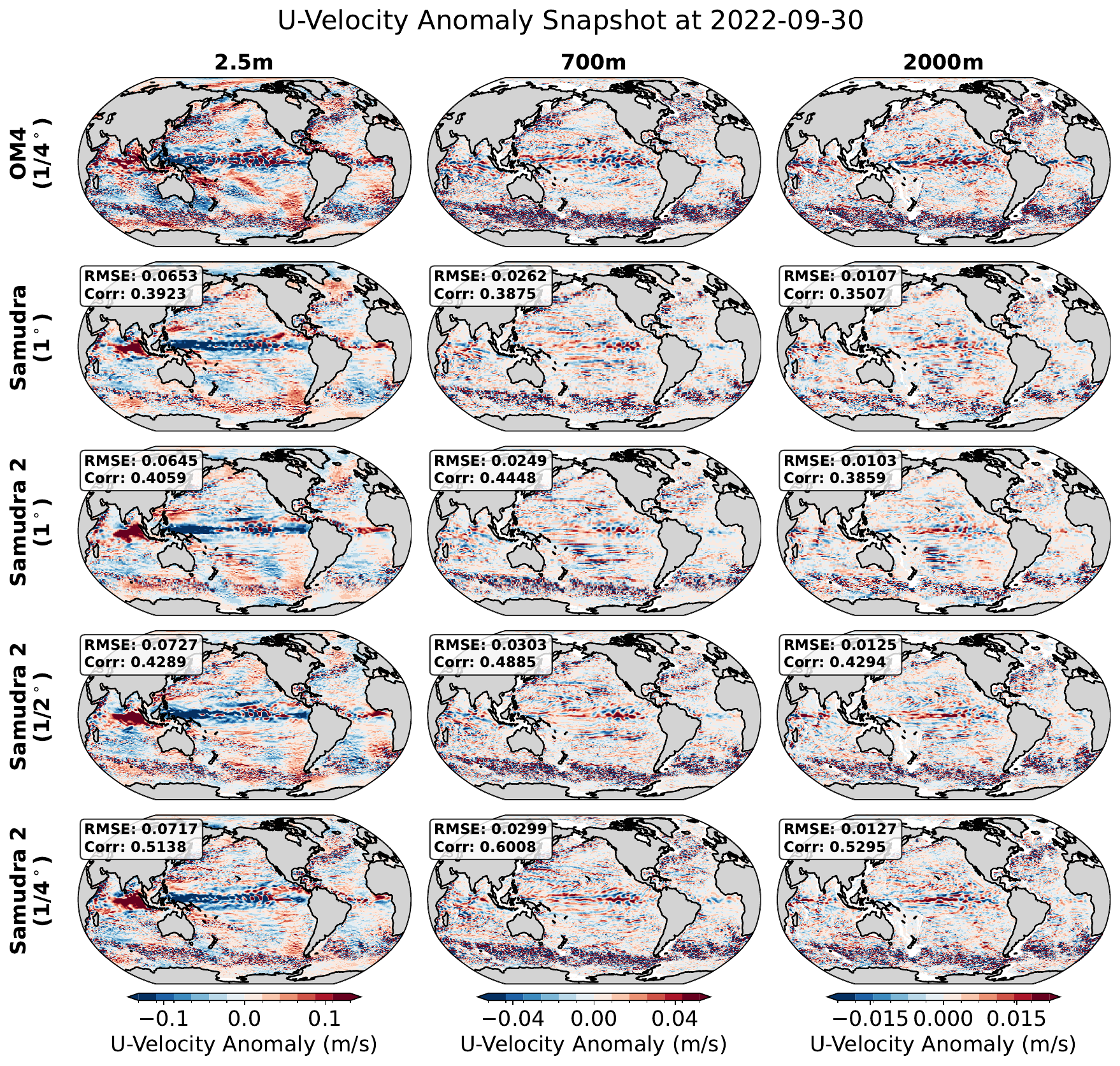}
\caption{Deseasonalized zonal velocity (u) anomaly snapshot at 2022-09-30, near the end of the 8-year rollout, at three depths (2.5\,m, 700\,m, 2000\,m) for OM4 ($1/4^\circ$), Samudra ($1^\circ$), and Samudra 2 at $1^\circ$, $1/2^\circ$, and $1/4^\circ$. RMSE and correlation values are annotated for each panel.}
\label{fig:snapshot_uo}
\end{figure}

\begin{figure}[htbp]
\centering
\includegraphics[width=\textwidth]{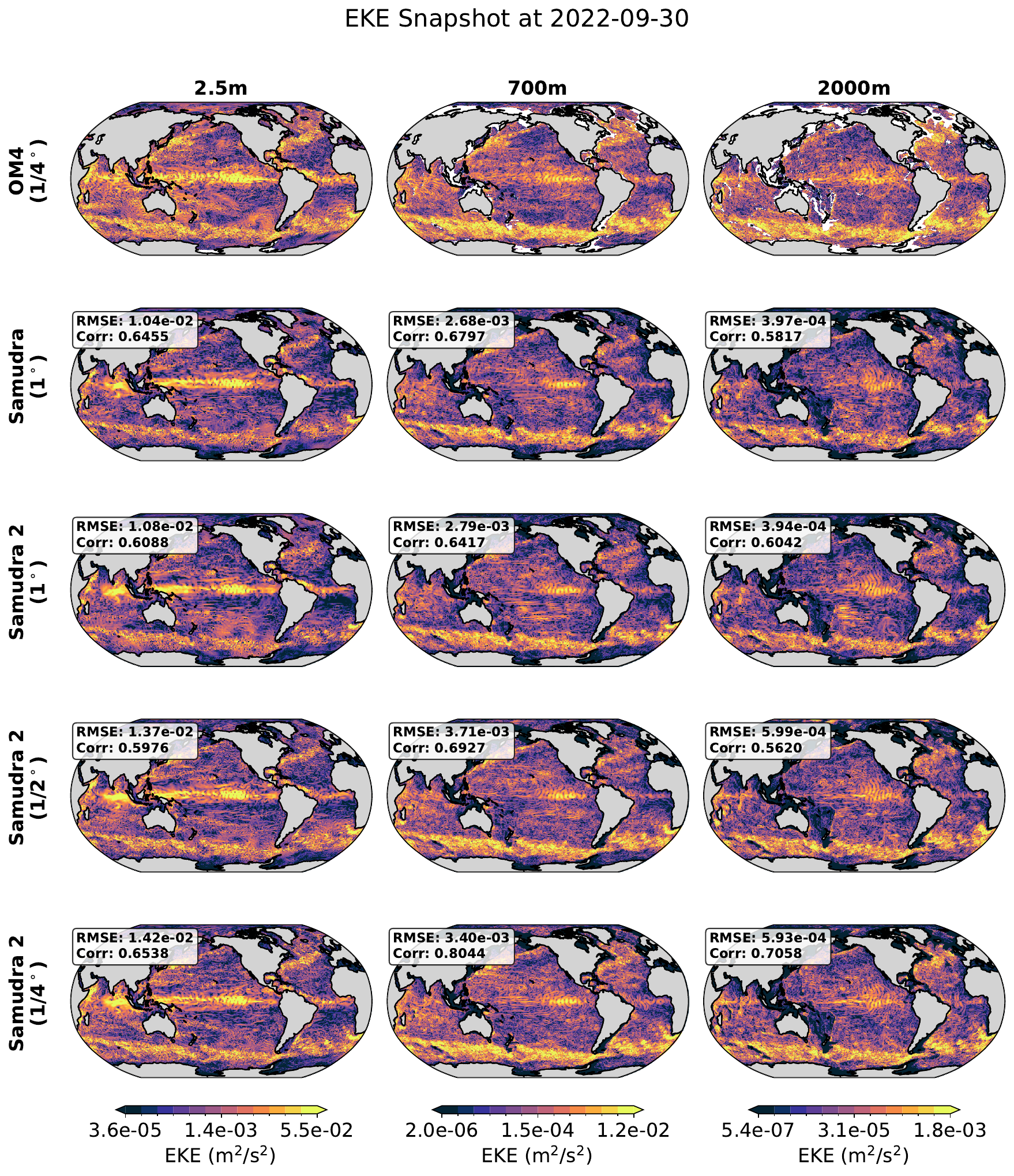}
\caption{Deseasonalized EKE snapshot at 2022-09-30, near the end of the 8-year rollout, at three depths (2.5\,m, 700\,m, 2000\,m) for OM4 ($1/4^\circ$), Samudra ($1^\circ$), and Samudra 2 at $1^\circ$, $1/2^\circ$, and $1/4^\circ$. RMSE and correlation values are annotated for each panel. Logarithmic color scale.}
\label{fig:snapshot_ke}
\end{figure}

The deseasonalized temperature anomaly snapshot is presented in the main text (Figure~\ref{fig:snapshot_thetao}).
Here we provide the corresponding snapshots for zonal velocity (u) and EKE (Figures~\ref{fig:snapshot_uo}--\ref{fig:snapshot_ke}).
Across all variables, correlation with the OM4 truth improves consistently with resolution, with the largest gains at 700\,m and 2000\,m: u-velocity correlation at 2000\,m increases from 0.39 ($1^\circ$) to 0.53 ($1/4^\circ$), and EKE from 0.60 to 0.71.
EKE snapshots (Figure~\ref{fig:snapshot_ke}) show a similar resolution dependence, with progressively finer spatial detail emerging at higher resolution.

\subsection{Spectral Analysis}
\label{sec:spectral}

\begin{figure}[htbp]
\centering
\includegraphics[width=\textwidth]{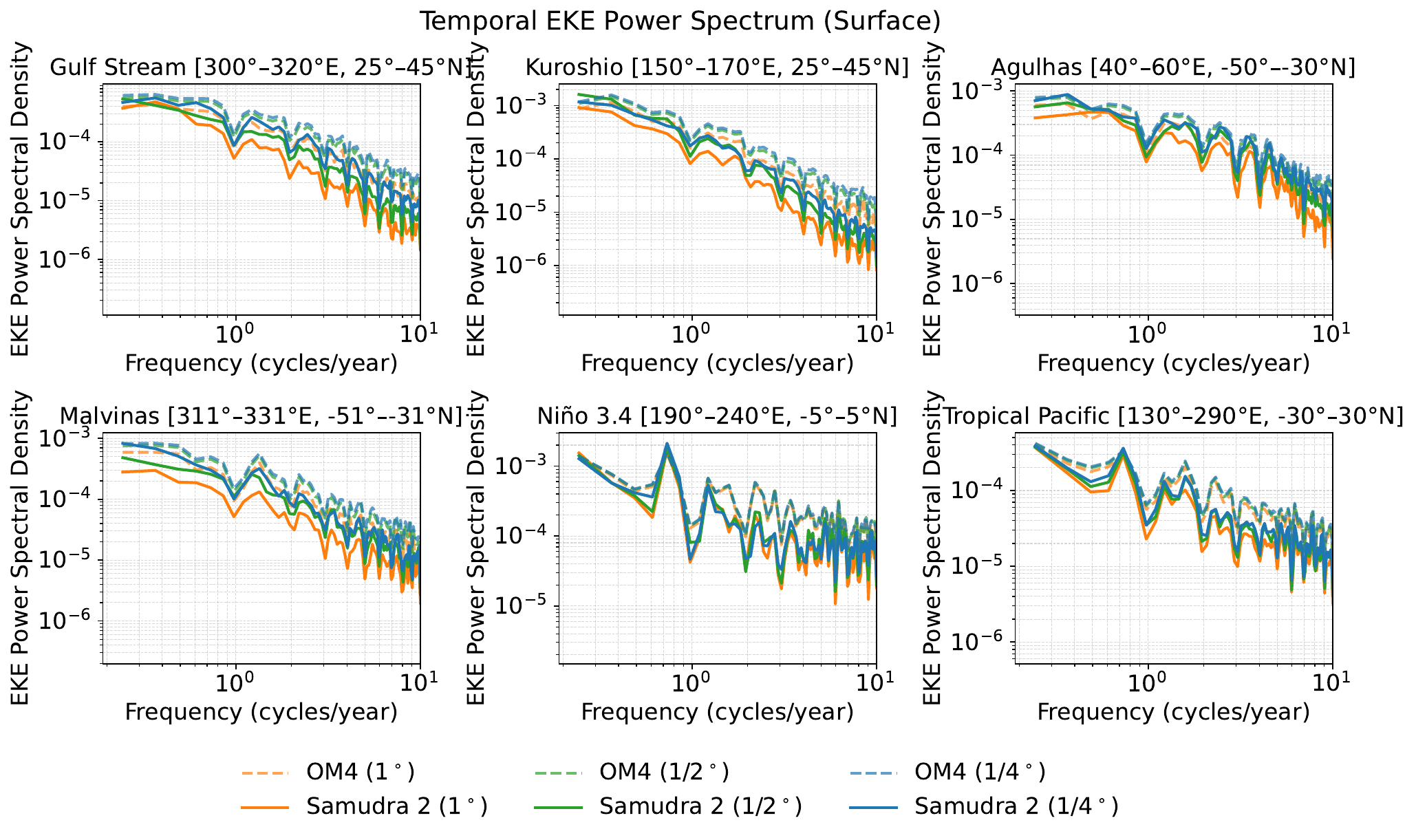}
\caption{Temporal EKE power spectra for six ocean regions. Each panel shows the OM4 truth (dashed) and Samudra 2 (solid) at three resolutions.}
\label{fig:spectrum_ke_temporal}
\end{figure}

\begin{figure}[htbp]
\centering
\includegraphics[width=\textwidth]{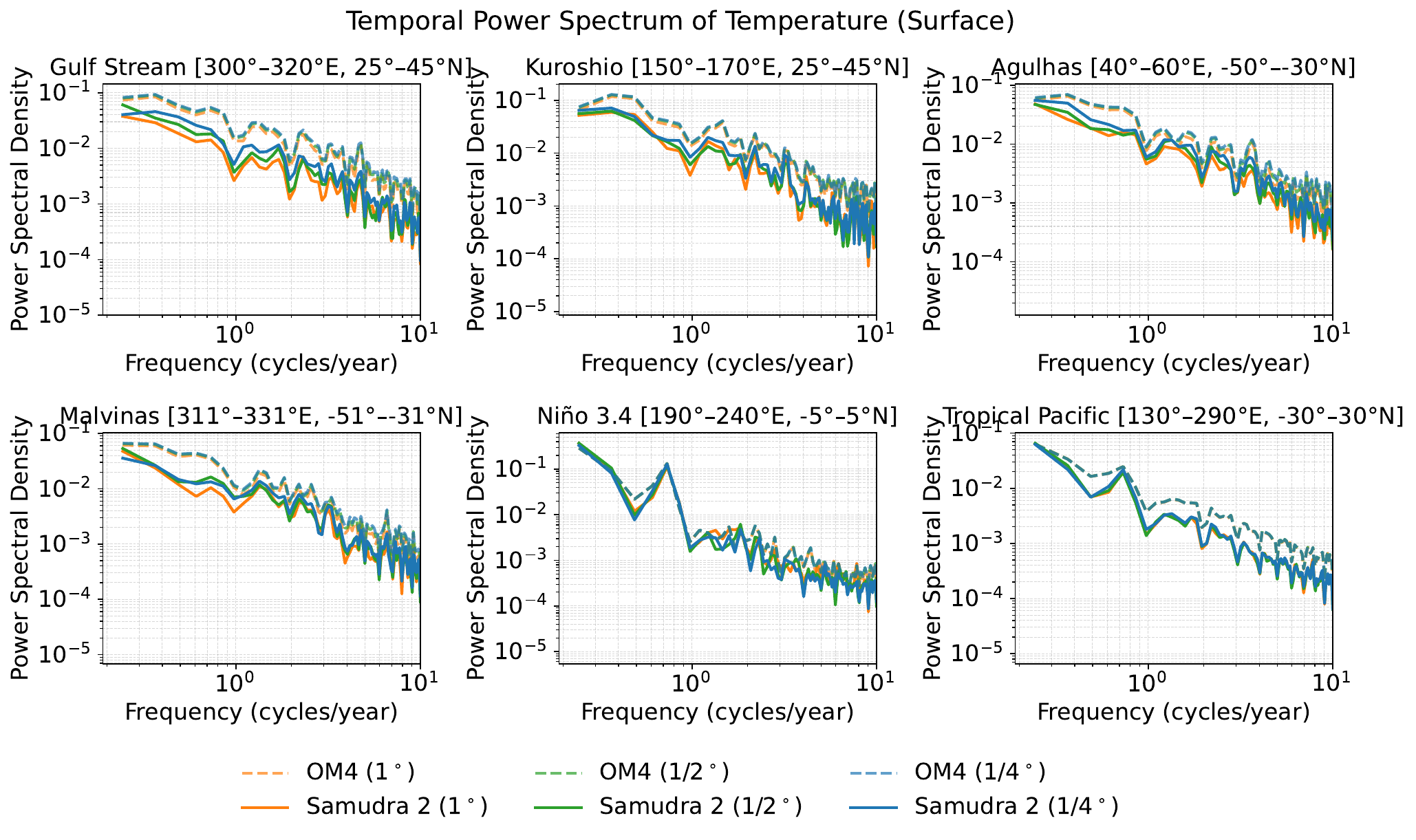}
\caption{Temporal temperature power spectra for six ocean regions. Each panel shows the OM4 truth (dashed) and Samudra 2 (solid) at three resolutions.}
\label{fig:spectrum_temporal}
\end{figure}

\begin{figure}[htbp]
\centering
\includegraphics[width=\textwidth]{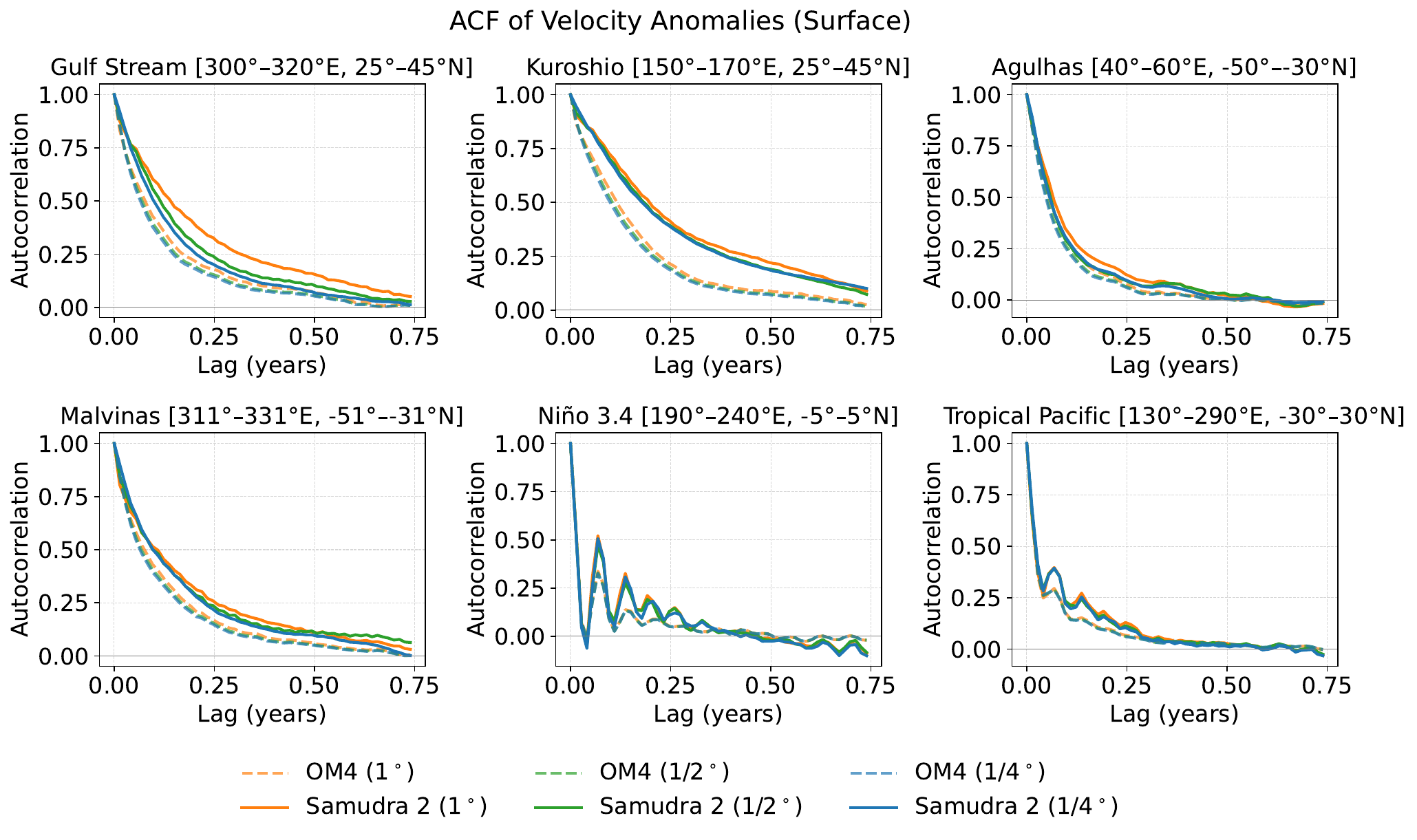}
\caption{Autocorrelation function (ACF) of surface velocity anomalies across six ocean regions. The Gulf Stream and Kuroshio regions show resolution-dependent slower decorrelation at higher resolution.}
\label{fig:acf_velocity}
\end{figure}

\begin{figure}[htbp]
\centering
\includegraphics[width=\textwidth]{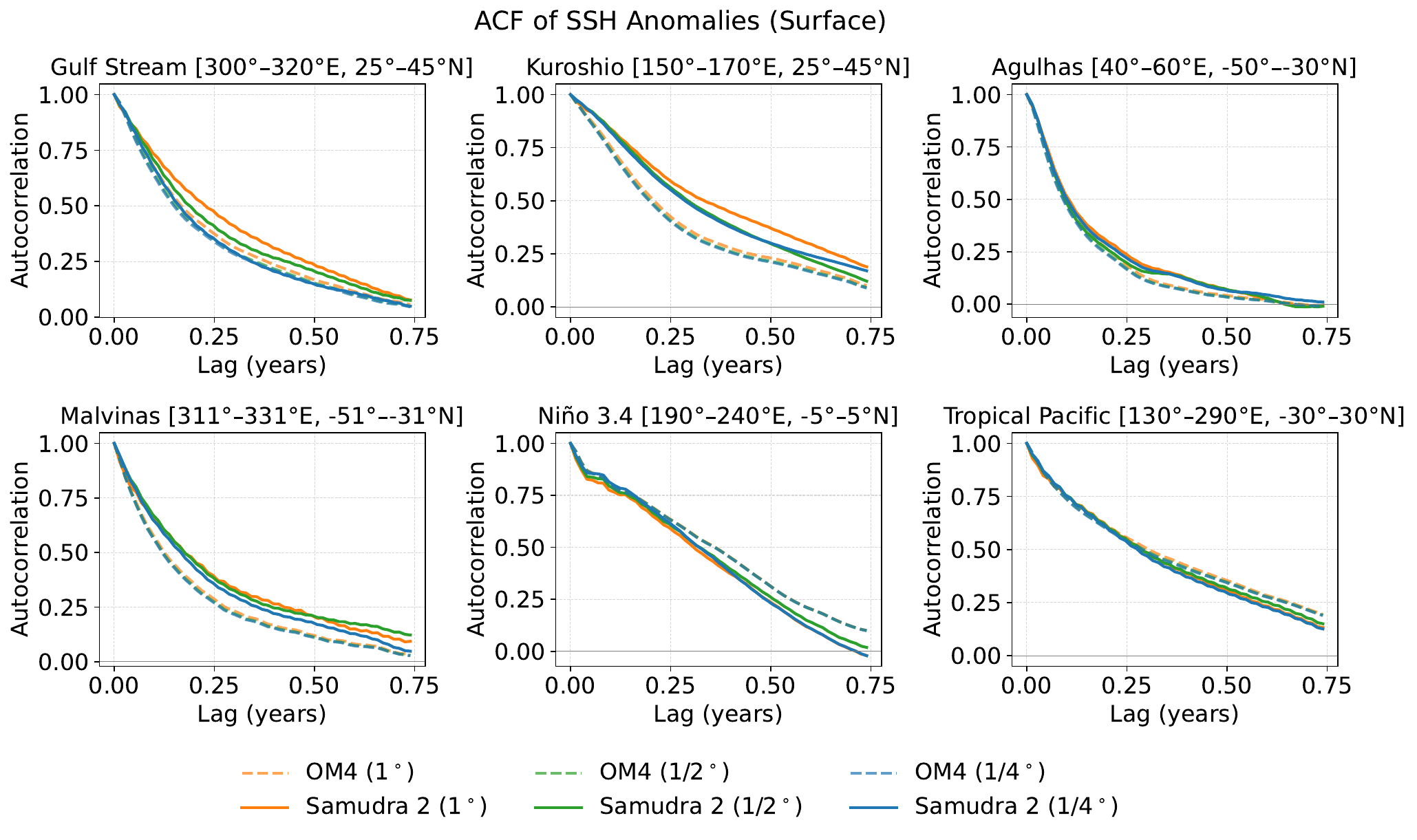}
\caption{Autocorrelation function of SSH anomalies across six ocean regions. All emulators closely match their respective OM4 truth.}
\label{fig:acf_ssh}
\end{figure}

The isotropic temperature and EKE spatial spectra are presented in the main text (Figure~\ref{fig:spectrum_combined}).
Here we provide additional temporal spectra and autocorrelation diagnostics.
The temporal EKE and temperature power spectra (Figures~\ref{fig:spectrum_ke_temporal}--\ref{fig:spectrum_temporal}) show close agreement at low frequencies and a growing deficit at higher frequencies.
The velocity and SSH autocorrelation functions (Figures~\ref{fig:acf_velocity}--\ref{fig:acf_ssh}) confirm that all emulators reproduce the decorrelation timescales of their respective OM4 truth, with the Gulf Stream and Kuroshio regions showing resolution-dependent slower decorrelation at higher resolution.

\subsection{Temporal Fidelity}
\label{sec:temporal}

\begin{figure}[htbp]
\centering
\includegraphics[width=0.82\textwidth]{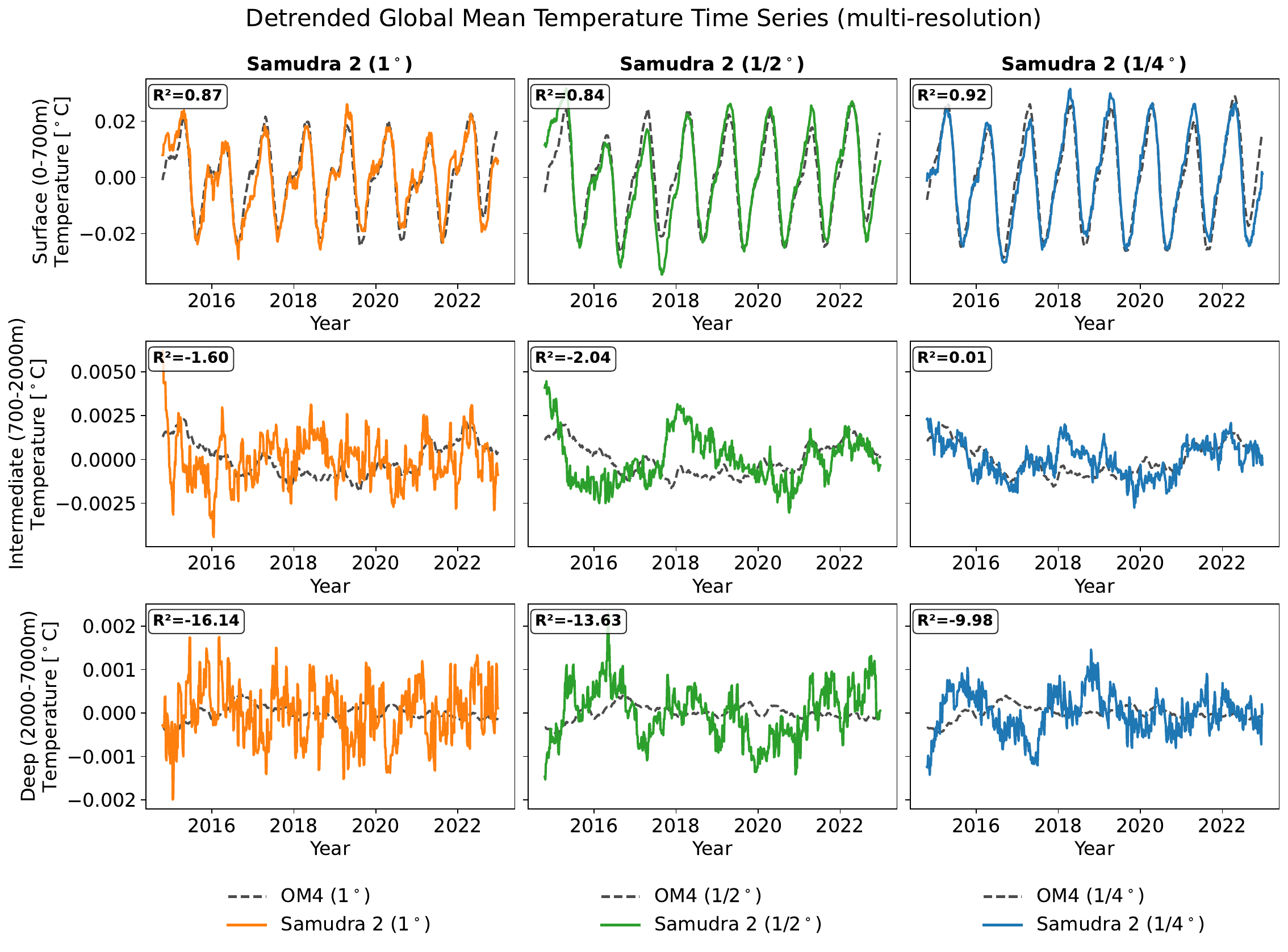}
\caption{Detrended global mean temperature time series for three depth ranges for Samudra 2 at three resolutions.}
\label{fig:timeseries_thetao}
\end{figure}

Figure~\ref{fig:timeseries_thetao} shows the detrended global mean temperature time series for Samudra 2 at all three resolutions, complementing the main-text discussion in Section~\ref{sec:variance}.
At deep levels (2000--7000\,m), $R^2$ remains negative at all resolutions due to the extremely small signal amplitude, but improves with resolution: the $1/4^\circ$ model achieves $R^2 = -9.98$ compared to $-16.14$ at $1^\circ$.

\begin{figure}[htbp]
\centering
\includegraphics[width=0.82\textwidth]{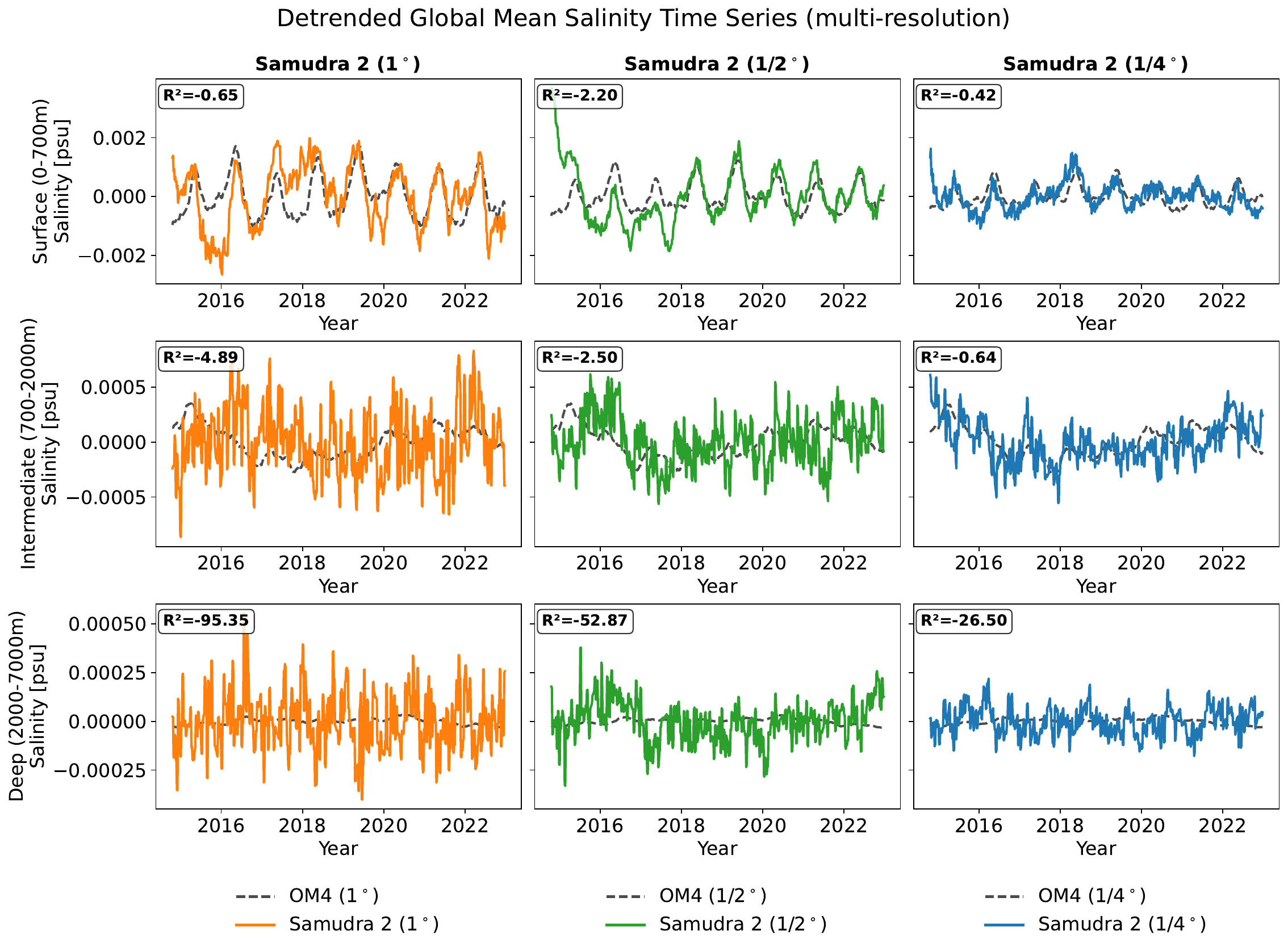}
\caption{Detrended global mean salinity time series for three depth ranges for Samudra 2 at three resolutions.}
\label{fig:timeseries_so}
\end{figure}

Figure~\ref{fig:timeseries_so} shows the corresponding detrended global mean salinity time series.
Salinity remains the most challenging variable: upper-ocean tracking is reasonable across resolutions, but intermediate and deep levels exhibit negative $R^2$ at all three resolutions, consistent with the variance metrics reported in the main text.
As with temperature, higher resolution yields modest improvements at depth, though deep-ocean salinity fidelity remains a key limitation.

\clearpage

\section{Supplementary Figures}
\label{sec:supp_figures}

\clearpage

\section{Supplementary Tables}
\label{sec:supp_tables}


\begin{table*}[h]
\centering
\footnotesize
\setlength{\tabcolsep}{3pt}
\caption{Temporal-variance evaluation for Salinity (ablation). All metrics are computed after removing the seasonal cycle from both predictions and the reference.}
\label{tab:temporal_variance_so}
\resizebox{\textwidth}{!}{%
\begin{tabular}{llcccccccccccc}
\toprule
\multirow{2}{*}{Region} & \multirow{2}{*}{Metric} & \multicolumn{4}{c}{Upper (0-700m)} & \multicolumn{4}{c}{Intermediate (700-2000m)} & \multicolumn{4}{c}{Deep (2000-7000m)} \\
\cmidrule(lr){3-6} \cmidrule(lr){7-10} \cmidrule(lr){11-14}
 &  & Samudra & Samudra-Wide & Samudra-DLoss & Samudra 2 & Samudra & Samudra-Wide & Samudra-DLoss & Samudra 2 & Samudra & Samudra-Wide & Samudra-DLoss & Samudra 2 \\
\midrule
Global & Var Corr & 0.256 & 0.801 & \best{0.938} & 0.844 & 0.429 & 0.420 & \best{0.473} & 0.467 & 0.171 & 0.242 & \best{0.298} & 0.269 \\
& Var RMSE & 0.419 & 0.219 & \best{0.178} & 0.211 & \best{5.79e-4} & 5.83e-4 & 5.91e-4 & 9.09e-4 & 5.49e-5 & 3.81e-5 & \best{2.13e-5} & 2.16e-5 \\
& Detrend RMSE & 0.0380 & 0.0374 & \best{0.0344} & 0.0345 & 0.00743 & 0.00803 & \best{0.00536} & 0.00549 & 0.00285 & 0.00280 & \best{0.00138} & 0.00143 \\
& Direct RMSE & 0.0413 & 0.0412 & 0.0380 & \best{0.0379} & 0.00785 & 0.00847 & \best{0.00594} & 0.00606 & 0.00291 & 0.00286 & \best{0.00147} & 0.00150 \\
\midrule
Gulf Stream & Var Corr & 0.879 & 0.850 & \best{0.880} & 0.753 & 0.477 & 0.317 & \best{0.481} & 0.314 & 0.191 & 0.233 & 0.174 & \best{0.265} \\
& Var RMSE & 0.00844 & 0.00877 & \best{0.00736} & 0.00814 & \best{6.65e-4} & 7.01e-4 & 0.00249 & 0.00499 & 2.59e-5 & 2.48e-5 & 7.85e-6 & \best{6.72e-6} \\
& Detrend RMSE & 0.0612 & \best{0.0603} & 0.0648 & 0.0672 & \best{0.0235} & 0.0248 & 0.0296 & 0.0349 & 0.00414 & 0.00403 & 0.00260 & \best{0.00257} \\
& Direct RMSE & 0.0686 & \best{0.0672} & 0.0691 & 0.0815 & \best{0.0260} & 0.0282 & 0.0351 & 0.0427 & 0.00476 & 0.00436 & 0.00343 & \best{0.00300} \\
\midrule
Kuroshio & Var Corr & 0.692 & 0.705 & 0.710 & \best{0.725} & 0.801 & 0.728 & \best{0.912} & 0.815 & 0.780 & 0.712 & \best{0.797} & 0.786 \\
& Var RMSE & 0.00191 & 0.00185 & 0.00204 & \best{0.00180} & 2.11e-4 & 2.45e-4 & \best{1.98e-4} & 2.09e-4 & 8.6e-6 & 7.82e-6 & \best{1.22e-6} & 2.09e-6 \\
& Detrend RMSE & 0.0439 & 0.0419 & \best{0.0393} & 0.0422 & 0.0132 & 0.0144 & \best{0.0114} & 0.0118 & 0.00247 & 0.00251 & \best{0.00113} & 0.00129 \\
& Direct RMSE & 0.0484 & 0.0459 & \best{0.0442} & 0.0458 & 0.0138 & 0.0148 & \best{0.0121} & 0.0123 & 0.00255 & 0.00256 & \best{0.00122} & 0.00133 \\
\midrule
Agulhas & Var Corr & 0.963 & 0.957 & \best{0.965} & 0.964 & 0.759 & 0.664 & \best{0.889} & 0.882 & 0.458 & 0.447 & \best{0.601} & 0.554 \\
& Var RMSE & 0.00247 & 0.00281 & \best{0.00215} & 0.00236 & 7.58e-5 & 1.66e-4 & 5.35e-5 & \best{5.31e-5} & 4.19e-5 & 3.3e-5 & \best{8.62e-6} & 8.95e-6 \\
& Detrend RMSE & 0.0402 & 0.0381 & \best{0.0359} & 0.0364 & 0.0117 & 0.0133 & \best{0.00808} & 0.00840 & 0.00554 & 0.00536 & \best{0.00282} & 0.00300 \\
& Direct RMSE & 0.0448 & 0.0461 & \best{0.0389} & 0.0402 & 0.0127 & 0.0144 & 0.0100 & \best{0.00969} & 0.00565 & 0.00551 & \best{0.00295} & 0.00310 \\
\midrule
Malvinas & Var Corr & 0.281 & 0.414 & 0.193 & \best{0.572} & 0.690 & 0.814 & \best{0.840} & 0.662 & 0.175 & 0.182 & \best{0.281} & 0.225 \\
& Var RMSE & 0.00226 & 0.00210 & 0.00244 & \best{0.00200} & 2.51e-4 & \best{2e-4} & 2.3e-4 & 2.95e-4 & 3.03e-5 & 3.1e-5 & \best{2.72e-5} & 2.77e-5 \\
& Detrend RMSE & 0.0445 & 0.0417 & 0.0430 & \best{0.0399} & 0.0157 & 0.0167 & \best{0.0131} & 0.0135 & 0.00533 & 0.00547 & \best{0.00391} & 0.00392 \\
& Direct RMSE & 0.0475 & 0.0462 & 0.0473 & \best{0.0456} & 0.0163 & 0.0172 & \best{0.0137} & 0.0143 & 0.00569 & 0.00588 & 0.00452 & \best{0.00434} \\
\midrule
Niño 3.4 & Var Corr & 0.644 & 0.425 & 0.739 & \best{0.789} & -0.478 & -0.529 & -0.480 & \best{-0.324} & 0.250 & 0.249 & 0.204 & \best{0.282} \\
& Var RMSE & 2.92e-4 & 4.14e-4 & 1.7e-4 & \best{1.57e-4} & 1.4e-4 & 1.78e-4 & 4.33e-5 & \best{3.75e-5} & 2.44e-5 & 2.97e-5 & 7.03e-6 & \best{5.67e-6} \\
& Detrend RMSE & 0.0249 & 0.0247 & 0.0199 & \best{0.0187} & 0.0115 & 0.0127 & 0.00641 & \best{0.00600} & 0.00463 & 0.00501 & 0.00238 & \best{0.00220} \\
& Direct RMSE & 0.0259 & 0.0264 & 0.0207 & \best{0.0198} & 0.0116 & 0.0128 & 0.00652 & \best{0.00607} & 0.00465 & 0.00503 & 0.00239 & \best{0.00222} \\
\midrule
Tropical Pacific & Var Corr & 0.200 & 0.709 & 0.804 & \best{0.816} & 0.231 & 0.227 & \best{0.356} & 0.352 & \best{0.202} & 0.183 & 0.165 & 0.145 \\
& Var RMSE & 0.788 & 0.0301 & \best{0.0261} & 0.0277 & 1.63e-4 & 1.73e-4 & 1.51e-4 & \best{1.5e-4} & 3.39e-5 & 3.05e-5 & \best{6.99e-6} & 7.2e-6 \\
& Detrend RMSE & 0.0337 & 0.0325 & 0.0308 & \best{0.0297} & 0.00694 & 0.00775 & 0.00435 & \best{0.00423} & 0.00311 & 0.00324 & \best{0.00153} & 0.00156 \\
& Direct RMSE & 0.0382 & 0.0368 & 0.0349 & \best{0.0331} & 0.00710 & 0.00794 & 0.00459 & \best{0.00442} & 0.00314 & 0.00328 & \best{0.00158} & 0.00160 \\
\bottomrule
\end{tabular}%
}
\end{table*}

\begin{table*}[h]
\centering
\footnotesize
\setlength{\tabcolsep}{3pt}
\caption{Temporal-variance evaluation for Kinetic Energy (ablation). All metrics are computed after removing the seasonal cycle from both predictions and the reference.}
\label{tab:temporal_variance_ke}
\resizebox{\textwidth}{!}{%
\begin{tabular}{llcccccccccccc}
\toprule
\multirow{2}{*}{Region} & \multirow{2}{*}{Metric} & \multicolumn{4}{c}{Upper (0-700m)} & \multicolumn{4}{c}{Intermediate (700-2000m)} & \multicolumn{4}{c}{Deep (2000-7000m)} \\
\cmidrule(lr){3-6} \cmidrule(lr){7-10} \cmidrule(lr){11-14}
 &  & Samudra & Samudra-Wide & Samudra-DLoss & Samudra 2 & Samudra & Samudra-Wide & Samudra-DLoss & Samudra 2 & Samudra & Samudra-Wide & Samudra-DLoss & Samudra 2 \\
\midrule
Global & Var Corr & 0.879 & 0.919 & \best{0.940} & 0.880 & 0.882 & \best{0.918} & 0.917 & 0.881 & 0.830 & 0.851 & \best{0.856} & 0.811 \\
& Var RMSE & 1.65e-4 & 1.49e-4 & \best{1.35e-4} & 1.73e-4 & 1.3e-5 & 1.12e-5 & \best{1.09e-5} & 1.38e-5 & 8.07e-7 & 7.91e-7 & \best{7.52e-7} & 8.35e-7 \\
& Detrend RMSE & 0.00217 & 0.00203 & \best{0.00196} & 0.00211 & 4.16e-4 & 3.89e-4 & \best{3.75e-4} & 4.04e-4 & 9.52e-5 & 9.14e-5 & \best{8.97e-5} & 9.38e-5 \\
& Direct RMSE & 0.00221 & 0.00208 & \best{0.00200} & 0.00216 & 4.25e-4 & 3.96e-4 & \best{3.84e-4} & 4.13e-4 & 9.65e-5 & 9.26e-5 & \best{9.1e-5} & 9.5e-5 \\
\midrule
Gulf Stream & Var Corr & 0.762 & 0.519 & \best{0.807} & 0.568 & \best{0.776} & 0.551 & 0.692 & 0.662 & 0.652 & 0.658 & \best{0.664} & 0.637 \\
& Var RMSE & \best{6.43e-5} & 8.27e-5 & 6.66e-5 & 7.66e-5 & \best{5.11e-7} & 5.44e-7 & 5.17e-7 & 5.23e-7 & \best{1.22e-7} & 1.26e-7 & 1.24e-7 & 1.27e-7 \\
& Detrend RMSE & 0.00330 & 0.00319 & \best{0.00305} & 0.00340 & 3.37e-4 & 3.33e-4 & \best{3.3e-4} & 3.42e-4 & 9.97e-5 & \best{9.59e-5} & 9.66e-5 & 9.73e-5 \\
& Direct RMSE & 0.00344 & 0.00336 & \best{0.00314} & 0.00358 & 3.5e-4 & 3.47e-4 & \best{3.43e-4} & 3.57e-4 & 1.02e-4 & \best{9.84e-5} & 9.91e-5 & 1e-4 \\
\midrule
Kuroshio & Var Corr & 0.814 & 0.431 & \best{0.872} & 0.748 & 0.778 & 0.374 & \best{0.864} & 0.809 & \best{0.674} & 0.601 & 0.621 & 0.618 \\
& Var RMSE & 1.12e-4 & 1.77e-4 & \best{9.75e-5} & 1.32e-4 & 1.61e-6 & 2.12e-6 & \best{1.39e-6} & 1.75e-6 & \best{5.11e-8} & 5.63e-8 & 5.51e-8 & 5.37e-8 \\
& Detrend RMSE & 0.00570 & 0.00549 & \best{0.00515} & 0.00546 & 4.7e-4 & 4.68e-4 & \best{4.53e-4} & 4.64e-4 & 1.03e-4 & 1e-4 & \best{9.94e-5} & 9.95e-5 \\
& Direct RMSE & 0.00597 & 0.00564 & \best{0.00536} & 0.00569 & 4.88e-4 & 4.8e-4 & \best{4.71e-4} & 4.82e-4 & 1.04e-4 & 1.01e-4 & 1.02e-4 & \best{1.01e-4} \\
\midrule
Agulhas & Var Corr & 0.964 & 0.971 & \best{0.978} & 0.959 & 0.966 & 0.969 & \best{0.977} & 0.964 & 0.956 & \best{0.959} & 0.946 & 0.939 \\
& Var RMSE & 1.54e-4 & \best{1.16e-4} & 1.32e-4 & 1.35e-4 & 7.99e-6 & \best{4.94e-6} & 6.94e-6 & 5.73e-6 & 6.6e-7 & \best{6.21e-7} & 7.2e-7 & 6.99e-7 \\
& Detrend RMSE & 0.00613 & 0.00556 & \best{0.00505} & 0.00601 & 0.00118 & 0.00107 & \best{9.85e-4} & 0.00117 & 2.52e-4 & 2.33e-4 & \best{2.3e-4} & 2.45e-4 \\
& Direct RMSE & 0.00620 & 0.00561 & \best{0.00513} & 0.00608 & 0.00119 & 0.00108 & \best{0.00100} & 0.00119 & 2.53e-4 & 2.34e-4 & \best{2.31e-4} & 2.47e-4 \\
\midrule
Malvinas & Var Corr & 0.521 & \best{0.753} & 0.749 & 0.568 & 0.479 & 0.711 & \best{0.739} & 0.550 & 0.646 & 0.819 & \best{0.824} & 0.687 \\
& Var RMSE & 4.36e-4 & 3.87e-4 & \best{3.85e-4} & 4.3e-4 & 6.27e-5 & 5.58e-5 & \best{5.47e-5} & 6.17e-5 & 2.09e-6 & 1.92e-6 & \best{1.9e-6} & 2.1e-6 \\
& Detrend RMSE & 0.00614 & 0.00555 & \best{0.00542} & 0.00599 & 0.00198 & 0.00181 & \best{0.00175} & 0.00193 & 5.2e-4 & 4.97e-4 & \best{4.76e-4} & 5.06e-4 \\
& Direct RMSE & 0.00628 & 0.00567 & \best{0.00554} & 0.00615 & 0.00202 & 0.00184 & \best{0.00180} & 0.00198 & 5.25e-4 & 5.03e-4 & \best{4.83e-4} & 5.1e-4 \\
\midrule
Niño 3.4 & Var Corr & 0.993 & 0.988 & \best{0.993} & 0.992 & 0.914 & \best{0.968} & 0.940 & 0.935 & \best{0.898} & 0.861 & 0.864 & 0.816 \\
& Var RMSE & 8.32e-5 & 9.18e-5 & \best{7.97e-5} & 9.43e-5 & 5.18e-7 & \best{1.9e-7} & 3.17e-7 & 3.77e-7 & \best{2.88e-8} & 3.3e-8 & 3.49e-8 & 3.86e-8 \\
& Detrend RMSE & 0.00742 & \best{0.00693} & 0.00702 & 0.00744 & 7.09e-4 & \best{6.34e-4} & 6.59e-4 & 6.65e-4 & 1.96e-4 & \best{1.83e-4} & 1.88e-4 & 1.9e-4 \\
& Direct RMSE & 0.00756 & \best{0.00702} & 0.00711 & 0.00756 & 7.16e-4 & \best{6.38e-4} & 6.67e-4 & 6.69e-4 & 1.97e-4 & \best{1.84e-4} & 1.89e-4 & 1.91e-4 \\
\midrule
Tropical Pacific & Var Corr & \best{0.970} & 0.953 & 0.966 & 0.959 & 0.658 & 0.683 & 0.679 & \best{0.705} & \best{0.948} & 0.907 & 0.932 & 0.728 \\
& Var RMSE & 3.11e-5 & 3.72e-5 & \best{2.95e-5} & 3.65e-5 & 9.51e-7 & 8.4e-7 & 9.34e-7 & \best{8.11e-7} & \best{3.38e-8} & 3.49e-8 & 3.95e-8 & 4.74e-8 \\
& Detrend RMSE & 0.00186 & 0.00179 & \best{0.00178} & 0.00185 & 2.26e-4 & \best{2.11e-4} & 2.24e-4 & 2.19e-4 & 6.06e-5 & \best{5.8e-5} & 5.84e-5 & 5.95e-5 \\
& Direct RMSE & 0.00189 & 0.00182 & \best{0.00182} & 0.00188 & 2.3e-4 & \best{2.14e-4} & 2.29e-4 & 2.23e-4 & 6.15e-5 & \best{5.88e-5} & 5.92e-5 & 6.04e-5 \\
\bottomrule
\end{tabular}%
}
\end{table*}

\end{document}